\documentclass[aps,prfluids,showkeys,superscriptaddress]{revtex4-2}
\usepackage[utf8]{inputenc} 
\usepackage[T1]{fontenc}    
\usepackage{url}           
\usepackage{booktabs}       
\usepackage{amsfonts}       
\usepackage{nicefrac}       
\usepackage{microtype}      
\usepackage{float}
\usepackage{caption}
\usepackage{caption}
\usepackage{soul}
\usepackage{comment}
\usepackage{enumitem}
\usepackage{subcaption}
\usepackage{soul}
\usepackage{amssymb,amsmath,amsthm}
\usepackage{changepage}
\usepackage{color,graphicx,float}
\usepackage{bm}
\usepackage{alltt,listings,tikz}
\usepackage[all]{xy}
\usepackage[left=0.8in,right=0.8in,top=1in,bottom=1in]{geometry}
\usepackage{hyperref}
\usepackage{fancyhdr}
\usepgflibrary{shadings}
\newcommand{\red}[1]{\ifmmode\mathbf{\textcolor{red}{#1}}\else \textbf{\textcolor{red}{#1}}\fi}

\hypersetup{
    colorlinks,
    linkcolor={blue},
    citecolor={blue},
    urlcolor={black}
}
\begin{document}

\title{\textbf{{\Large Modeling film flows down a rotating slippery cylinder}}}

\author{Souradip Chattopadhyay}
\email{schatto5@ncsu.edu (ORCID: 0000-0002-4418-6201)}
\affiliation{Department of Mathematics, North Carolina State University, Raleigh, North Carolina 27695, USA}

\author{Amar K. Gaonkar}
\email{amar.gaonkar@iitdh.ac.in}
\affiliation{Department of Mechanical, Materials and Aerospace Engineering, IIT Dharwad, Karnataka 580011, India}

\author{Hangjie Ji}
\email{hangjie\_ji@ncsu.edu (corresponding author)}
\affiliation{Department of Mathematics, North Carolina State University, Raleigh, North Carolina 27695, USA}

\begin{abstract}
This study investigates the nonlinear stability and dynamics of gravity-driven viscous films on a vertical rotating cylinder, considering both outer and inner surface flows with slip conditions at the cylinder wall. We develop an asymptotic model for the combined effects of rotation and wall slippage. Linear stability analysis indicates that wall slippage enhances instability on both surfaces, while rotation has differing impacts: it amplifies instability due to slip for outer surface flow but reduces it for inner surface flow. A weakly nonlinear stability analysis is then conducted to explore the combined impact of rotation and wall slip on flow stability beyond the linear regime, including the bifurcation of the nonlinear evolution equation for both surfaces. The traveling wave solution of the model is analyzed, showing how rotation affects nonlinear wave speed with a slippery wall. A stability analysis of the traveling wave solutions is also performed. Numerical simulations of the nonlinear evolution of the free surface reveal that increasing slip length enhances the choke phenomenon in inner surface flow, while rotation can delay this effect. Additionally, simulations show that for flow along the outer surface of a slippery rotating cylinder, the film tends to break up into droplets in the presence of rotation.
\end{abstract}

\maketitle
 
\section{Introduction}\label{sec:1}
Coating non-flat objects is crucial for manufacturing various products, such as aerospace components \cite{garg1989} and three-dimensional printed parts \cite{zhu2015}, to protect their surfaces. Understanding the behavior of a thin liquid film flowing down a vertical cylinder's inner or outer surface provides valuable insights into this process. This investigation reveals complex interfacial dynamics, including droplet formation and traveling wave patterns \cite{quere1990,kalliadasis1994}, which are critical for applications like coating insulation on wires \cite{quere1999}, gas absorption \cite{grunig2012}, desalination \cite{sadeghpour2019}, and surface patterning on cylindrical substrates \cite{gau1999}. The flow on the inner surface of cylindrical geometries, such as within tubes, is also important in applications such as lung airways \cite{grotberg1994}.

\par When a thin liquid film is applied to the surface of a fiber or cylinder, it typically experiences capillary instability \cite{craster2006,cm2008,kalliadasis1994}. Theoretical and experimental studies have examined the dynamics of liquid film flow along a vertical cylindrical geometry \cite{kalliadasis1994, quere1990}, providing a detailed understanding of the underlying physics. Unlike films on flat substrates, those on cylindrical surfaces exhibit inherent instability due to the additional azimuthal curvature of the free surface \cite{quere1990}, resulting from the interaction of two distinct instability mechanisms. The first is the Rayleigh–Plateau (RP) instability, influenced by the gravity-driven flow in the axisymmetric film coating a cylindrical fiber. The second is the hydrodynamic instability of a falling film due to inertia, known as the Kapitza mode of instability \cite{duprat2009}. The challenge becomes more pronounced when the cylinder rotates around its vertical axis, as gravitational, inertial, viscous, surface tension and centrifugal forces interact, potentially leading to instabilities and coating failures. When the cylinder is horizontal, various studies have reported that rotating cylinders effectively control coating flow stability and dynamics, which is crucial in numerous industrial applications for surface control and protection \cite{ruschak1976,ashmore2003,hosoi1999}. Experimental investigations by Thoroddsen \& Mahadevan \cite{thoroddsen1997}, as well as Kozlov \& Polezhaev \cite{kozlov2015}, have demonstrated how rotation influences flow dynamics in case of horizontal scenarios. Compared to horizontal cylinders, there are limited studies on liquid film flow along a rotating vertical cylinder or fiber in the current literature. Experimental results from rotating horizontal cylinders inspire us to focus on existing models of vertical cylinders under rotation, examining flow along the outer and inner surfaces. By extending these studies, we aim to understand the underlying physics better.

\par Frenkel \cite{frenkel1992} examined the case of a large cylinder radius (resulting in a thin film) and derived a Benney-type equation for the interface. However, their model was not derived asymptotically, and the bead velocity was overestimated.  Kliakhandler et al. \cite{kliakhandler2001} and Craster \& Matar \cite{craster2006} investigated the instability of a uniform film at higher flow rates by regularly wetting the fiber from the top. They reported the nonlinear dynamics of axisymmetric waves occurring far from the inlet. For a thin liquid film on the outer surface of a vertical cylindrical fiber, Kim et al. \cite{kim2024} studied a positivity-preserving finite difference method. A recent study by Taranets et al. \cite{taranets2024} proved the existence of weak solutions for the flow of a thin liquid film outside a vertical fiber and established the presence of a traveling wave solution for the system. Novbari \& Oron \cite{novbari2009} employed the energy integral method to investigate the nonlinear dynamics of a thin film flowing on the outer surface of a vertical cylinder. Various models have been reported to understand complex dynamical phenomena of liquid film flow along a vertical fiber/cylinder \cite{oron1997,craster2009}. Typical long-wave models either assume that the film thickness is much smaller than the radius of the cylinder \cite{kalliadasis1994,halpern2017} or that the film radius is much smaller than its characteristic length \cite{craster2006,liu2014}. These models simplify the Navier-Stokes equations into single evolution equations for the film thickness, making them suitable for analyzing thin-film flows with low Reynolds numbers. More advanced models, such as the integral-boundary-layer model \cite{sisoev2006} and the weighted-residual integral-boundary-layer model \cite{cm2008}, have been developed for scenarios involving a moderate Reynolds number. These models formulate evolution equations for film thickness and volumetric flow rates. 

\par The above-mentioned models assumed the conventional no-slip boundary condition at the cylinder/fiber wall interface. However, numerous studies have explored the potential role of slip at the solid-liquid interface, which can enhance free-surface instability. Haefner et al. \cite{haefner2015} investigated the influence of slip on the RP instability in a film along a fiber. Their study, combining experiment and theory, highlighted that wall slip significantly affects the growth rate of instabilities. Applying lubrication theory, the model proposed by Halpern \& Wei \cite{halpern2017} concluded that slip length enhances drop formation along a vertical fiber. Ji et al. \cite{ji2019} presented a study combining experimental and numerical results on thin film flows along the outer surface of a slippery fiber. Their lubrication model incorporated slip boundary conditions, nonlinear curvature terms, and a film stabilization term. Chao et al. \cite{chao2018} explored the combined influence of wall slip and thermocapillary effects on thin film flow along a vertical cylinder, demonstrating that both factors enhance the RP instability. Chattopadhyay \cite{souradip2023aaaa} extended this investigation to include the impact of a chemical reaction. His results indicated that an endothermic reaction intensifies the RP instability in the presence of a slippery wall compared to the no-slip case, whereas an exothermic reaction reduces it. The studies  \cite{haefner2015,halpern2017,ji2019,chao2018,souradip2023aaaa} examined the influence of slip length on the stability and dynamics of thin film flow along the outer surface of a slippery cylinder/fiber. A recent study by Schwitzerlett et al. \cite{schwitzerlett2023} developed an asymptotic model examining flow within a vertical tube with a slippery wall. Their findings revealed that wall slippage destabilizes flow instability, similar to what is observed on the outer surface of a vertical cylinder/fiber. Additionally, they concluded that the presence of slip length enhances the formation of plugs.

\par Although numerous studies have examined the dynamics and stability of vertical cylindrical geometry, very few have addressed the effect of cylinder rotation. D{\'a}valos-Orozco \& Ruiz-Chavarria \cite{davalos1993} investigated the linear stability of a fluid layer flowing inside and outside of a rotating vertical cylinder. Their analysis incorporated two additional approximations: small wavenumber and small Reynolds number. Using a long-wave expansion method, Chen et al. \cite{chen2004} developed a model to examine the stability and dynamics of a condensate liquid film along the outer surface of a rotating cylinder. They performed both linear and weakly nonlinear stability analyses. They found that the rotation increases the instability of the flow. Rietz et al. \cite{riestz2017} conducted an experimental investigation into developing a thin film on the outer surface of a vertical rotating cylinder. Their study specifically addressed the generation of 2D and 3D waves, the formation of rivulets, and dripping phenomena. Liu \& Ding \cite{liu2020} examined the dynamics of a coating flow over a fiber that rotates about its axis. They formulated an evolution equation for the surface using the long-wave theory. Their findings indicated that rotation has a destabilizing effect. Furthermore, their spatiotemporal stability analysis revealed that rotation enhances absolute instability. Farooq et al. \cite{farooq2020} investigated the dynamics of a falling film on the inner surface of a rotating cylinder. They utilized numerical simulations with the VOF method to explore the effects of rotation on interfacial dynamics in two and three dimensions. Their findings highlight rotation's significant influence on interfacial waves' structure. Mukhopadhyay et al. \cite{souradip2020} examined how the dynamics of thin liquid films behave on the outer surface of a vertical rotating cylinder under non-isothermal conditions. They found that rotation intensifies flow instability when thermal effects are present.

\par In this study, we examine the flow of a thin film over the surface of a vertical cylinder under gravity, where the cylinder rotates about its axis. The primary motivation is to develop a mathematical model that explores two distinct scenarios: one where the film flows along the outer surface of the cylinder and another where it flows along the inner surface. Previous studies by Ji et al. \cite{ji2019}, Chao et al. \cite{chao2018}, and Schwitzerlett et al. \cite{schwitzerlett2023} have shown that wall slippage enhances flow instability in similar configurations. This work focuses on how cylinder rotation impacts flow dynamics and stability exacerbated by wall slippage on both outer and inner surfaces.  We extend the stability analysis beyond the linear regime, focusing on bifurcation scenarios involving the two key parameters: wall slip and rotation. Additionally, we investigate the influence of these parameters on the propagation speed of traveling wave solutions. Further contributions include a comprehensive stability assessment of these traveling waves for both outer and inner surface flows. Finally, we explore the effect of cylinder rotation on the breakup dynamics of liquid flows along the outer surface and on the plug formation in flows along the inner surface, a phenomenon reported by Liu \& Ding \cite{liu2017} and Schwitzerlett et al. \cite{schwitzerlett2023}. Our study aims to provide insights into delaying such plug formation under varying rotation and slip length conditions. 

\par We organize the present paper as follows. Section \ref{sec:2} introduces a model incorporating wall slip and rotation for a rotating vertical cylinder. Section \ref{sec:3} presents a linear stability analysis for the flow. Section \ref{sec:4} presents a weakly nonlinear stability analysis, exploring the combined effects of wall slip and rotation beyond the linear regime. Section \ref{sec:5} focuses on investigating traveling wave solutions and their stability in the presence of slip and rotation. In Section \ref{sec:6}, we perform nonlinear simulations of the evolution equation. We discuss the influence of rotation on the choke behavior when the flow is along the inner surface of the slippery cylinder and on the breakup behavior when the flow is along the outer surface of the slippery cylinder. Finally, Section \ref{sec:7} summarizes the study's key findings.

\section{Model}\label{sec:2}
\begin{figure} 
\tikzset {_ajzn820lw/.code = {\pgfsetadditionalshadetransform{ \pgftransformshift{\pgfpoint{0 bp } { 0 bp }  }  \pgftransformrotate{0 }  \pgftransformscale{2 }  }}}
\pgfdeclarehorizontalshading{_f17kgnf6r}{150bp}{rgb(0bp)=(1,1,1);
rgb(37.5bp)=(1,1,1);
rgb(50bp)=(0.95,0.95,0.95);
rgb(50.25bp)=(0.93,0.93,0.93);
rgb(62.5bp)=(1,1,1);
rgb(100bp)=(1,1,1)}
\tikzset {_ms4xpsc8s/.code = {\pgfsetadditionalshadetransform{ \pgftransformshift{\pgfpoint{0 bp } { 0 bp }  }  \pgftransformrotate{0 }  \pgftransformscale{2 }  }}}
\pgfdeclarehorizontalshading{_lyr0o4cx2}{150bp}{rgb(0bp)=(1,1,1);
rgb(37.5bp)=(1,1,1);
rgb(50bp)=(0.95,0.95,0.95);
rgb(50.25bp)=(0.93,0.93,0.93);
rgb(62.5bp)=(1,1,1);
rgb(100bp)=(1,1,1)}
\tikzset{every picture/.style={line width=0.75pt}} 
\begin{tikzpicture}[x=0.4pt,y=0.4pt,yscale=-1,xscale=1]

\draw   (243,87.05) -- (243,392.95) .. controls (243,401.81) and (219.05,409) .. (189.5,409) .. controls (159.95,409) and (136,401.81) .. (136,392.95) -- (136,87.05) .. controls (136,78.19) and (159.95,71) .. (189.5,71) .. controls (219.05,71) and (243,78.19) .. (243,87.05) .. controls (243,95.91) and (219.05,103.1) .. (189.5,103.1) .. controls (159.95,103.1) and (136,95.91) .. (136,87.05) ;

\path  [shading=_f17kgnf6r,_ajzn820lw] (136,87.05) -- (243,87.05) -- (243,416) -- (136,416) -- cycle ; 
 \draw   (136,87.05) -- (243,87.05) -- (243,416) -- (136,416) -- cycle ; 

\draw [line width=0.75]    (188.5,38.53) -- (188.5,447.53) ;
\draw [shift={(188.5,450.53)}, rotate = 270] [fill={rgb, 255:red, 0; green, 0; blue, 0 }  ][line width=0.08]  [draw opacity=0] (10.72,-5.15) -- (0,0) -- (10.72,5.15) -- (7.12,0) -- cycle    ;

\draw [color={rgb, 255:red, 255; green, 255; blue, 255 }  ,draw opacity=1 ] [dash pattern={on 4.5pt off 4.5pt}]  (262,403) -- (369,403) ;

\draw    (190,165) -- (239,165) ;
\draw [shift={(242,165)}, rotate = 180] [fill={rgb, 255:red, 0; green, 0; blue, 0 }  ][line width=0.08]  [draw opacity=0] (8.93,-4.29) -- (0,0) -- (8.93,4.29) -- cycle    ;
 
\draw    (244,221) -- (289,221) ;
\draw [shift={(292,221)}, rotate = 180] [fill={rgb, 255:red, 0; green, 0; blue, 0 }  ][line width=0.08]  [draw opacity=0] (8.93,-4.29) -- (0,0) -- (8.93,4.29) -- cycle    ;
 
\draw    (242,165) -- (193,165) ;
\draw [shift={(190,165)}, rotate = 360] [fill={rgb, 255:red, 0; green, 0; blue, 0 }  ][line width=0.08]  [draw opacity=0] (8.93,-4.29) -- (0,0) -- (8.93,4.29) -- cycle    ;

\draw    (348,317) -- (348,328) -- (348,414) ;
\draw [shift={(348,417)}, rotate = 270] [fill={rgb, 255:red, 0; green, 0; blue, 0 }  ][line width=0.08]  [draw opacity=0] (10.72,-5.15) -- (0,0) -- (10.72,5.15) -- (7.12,0) -- cycle    ;

\draw    (125,87) -- (307,87) ;
\draw [shift={(310,87)}, rotate = 180] [fill={rgb, 255:red, 0; green, 0; blue, 0 }  ][line width=0.08]  [draw opacity=0] (10.72,-5.15) -- (0,0) -- (10.72,5.15) -- (7.12,0) -- cycle    ;

\draw   (542,87.05) -- (542,392.95) .. controls (542,401.81) and (518.05,409) .. (488.5,409) .. controls (458.95,409) and (435,401.81) .. (435,392.95) -- (435,87.05) .. controls (435,78.19) and (458.95,71) .. (488.5,71) .. controls (518.05,71) and (542,78.19) .. (542,87.05) .. controls (542,95.91) and (518.05,103.1) .. (488.5,103.1) .. controls (458.95,103.1) and (435,95.91) .. (435,87.05) ;

\path  [shading=_lyr0o4cx2,_ms4xpsc8s] (435,87.05) -- (542,87.05) -- (542,416) -- (435,416) -- cycle ; 
 \draw   (435,87.05) -- (542,87.05) -- (542,416) -- (435,416) -- cycle ; 

\draw [line width=0.75]    (489.5,38.53) -- (489.5,447.53) ;
\draw [shift={(489.5,450.53)}, rotate = 270] [fill={rgb, 255:red, 0; green, 0; blue, 0 }  ][line width=0.08]  [draw opacity=0] (10.72,-5.15) -- (0,0) -- (10.72,5.15) -- (7.12,0) -- cycle    ;

\draw    (414,87) -- (596,87) ;
\draw [shift={(599,87)}, rotate = 180] [fill={rgb, 255:red, 0; green, 0; blue, 0 }  ][line width=0.08]  [draw opacity=0] (10.72,-5.15) -- (0,0) -- (10.72,5.15) -- (7.12,0) -- cycle    ;

\draw [color={rgb, 255:red, 74; green, 144; blue, 226 }  ,draw opacity=1 ][line width=1.5]    (467,195) .. controls (450,184) and (443,172) .. (447,87) ;

\draw [color={rgb, 255:red, 74; green, 144; blue, 226 }  ,draw opacity=1 ][line width=1.5]    (451,277) .. controls (428,249) and (515,232) .. (467,195) ;

\draw [color={rgb, 255:red, 74; green, 144; blue, 226 }  ,draw opacity=1 ][line width=1.5]    (453,360) .. controls (436,332) and (521,335) .. (451,277) ;
 
\draw [color={rgb, 255:red, 74; green, 144; blue, 226 }  ,draw opacity=1 ][line width=1.5]    (451,431) .. controls (429,395) and (506,423) .. (453,360) ;

\draw [color={rgb, 255:red, 74; green, 144; blue, 226 }  ,draw opacity=1 ][line width=1.5]    (521,283) .. controls (564,252) and (482,244) .. (508,205) ;

\draw [color={rgb, 255:red, 74; green, 144; blue, 226 }  ,draw opacity=1 ][line width=1.5]    (519,373) .. controls (561,329) and (464,344) .. (521,283) ;

\draw [color={rgb, 255:red, 74; green, 144; blue, 226 }  ,draw opacity=1 ][line width=1.5]    (525,429) .. controls (545,393) and (483,432) .. (519,373) ;

\draw [color={rgb, 255:red, 74; green, 144; blue, 226 }  ,draw opacity=1 ][line width=1.5]    (508,205) .. controls (545,178) and (526,169) .. (532,87) ;

\draw [color={rgb, 255:red, 74; green, 144; blue, 226 }  ,draw opacity=1 ][line width=1.5]    (278,197) .. controls (261,186) and (254,174) .. (258,89) ;

\draw [color={rgb, 255:red, 74; green, 144; blue, 226 }  ,draw opacity=1 ][line width=1.5]    (262,279) .. controls (239,251) and (326,234) .. (278,197) ;

\draw [color={rgb, 255:red, 74; green, 144; blue, 226 }  ,draw opacity=1 ][line width=1.5]    (264,362) .. controls (247,334) and (332,337) .. (262,279) ;

\draw [color={rgb, 255:red, 74; green, 144; blue, 226 }  ,draw opacity=1 ][line width=1.5]    (262,433) .. controls (240,397) and (317,425) .. (264,362) ;

\draw [color={rgb, 255:red, 74; green, 144; blue, 226 }  ,draw opacity=1 ][line width=1.5]    (101,205) .. controls (138,178) and (119,169) .. (125,87) ;

\draw [color={rgb, 255:red, 74; green, 144; blue, 226 }  ,draw opacity=1 ][line width=1.5]    (114,283) .. controls (157,252) and (75,244) .. (101,205) ;

\draw [color={rgb, 255:red, 74; green, 144; blue, 226 }  ,draw opacity=1 ][line width=1.5]    (112,373) .. controls (154,329) and (57,344) .. (114,283) ;
 
\draw [color={rgb, 255:red, 74; green, 144; blue, 226 }  ,draw opacity=1 ][line width=1.5]    (118,429) .. controls (138,393) and (76,432) .. (112,373) ;

\draw    (436,217) -- (478,217) ;
\draw [shift={(481,217)}, rotate = 180] [fill={rgb, 255:red, 0; green, 0; blue, 0 }  ][line width=0.08]  [draw opacity=0] (8.93,-4.29) -- (0,0) -- (8.93,4.29) -- cycle    ;

\draw    (481,217) -- (439,217) ;
\draw [shift={(436,217)}, rotate = 360] [fill={rgb, 255:red, 0; green, 0; blue, 0 }  ][line width=0.08]  [draw opacity=0] (8.93,-4.29) -- (0,0) -- (8.93,4.29) -- cycle    ;

\draw    (289,221) -- (247,221) ;
\draw [shift={(244,221)}, rotate = 360] [fill={rgb, 255:red, 0; green, 0; blue, 0 }  ][line width=0.08]  [draw opacity=0] (8.93,-4.29) -- (0,0) -- (8.93,4.29) -- cycle    ;

\draw    (360,172) -- (268.27,220.07) ;
\draw [shift={(266.5,221)}, rotate = 332.34] [fill={rgb, 255:red, 0; green, 0; blue, 0 }  ][line width=0.08]  [draw opacity=0] (12,-3) -- (0,0) -- (12,3) -- cycle    ;

\draw    (367,172) -- (456.71,216.12) ;
\draw [shift={(458.5,217)}, rotate = 206.19] [fill={rgb, 255:red, 0; green, 0; blue, 0 }  ][line width=0.08]  [draw opacity=0] (12,-3) -- (0,0) -- (12,3) -- cycle    ;

\draw    (487,131) -- (438,131) ;
\draw [shift={(435,131)}, rotate = 360] [fill={rgb, 255:red, 0; green, 0; blue, 0 }  ][line width=0.08]  [draw opacity=0] (8.93,-4.29) -- (0,0) -- (8.93,4.29) -- cycle    ;

\draw    (435,131) -- (484,131) ;
\draw [shift={(487,131)}, rotate = 180] [fill={rgb, 255:red, 0; green, 0; blue, 0 }  ][line width=0.08]  [draw opacity=0] (8.93,-4.29) -- (0,0) -- (8.93,4.29) -- cycle    ;
 
\draw    (166,48) .. controls (176.67,64.49) and (194.87,71.57) .. (213.29,48.25) ;
\draw [shift={(215,46)}, rotate = 126.16] [fill={rgb, 255:red, 0; green, 0; blue, 0 }  ][line width=0.08]  [draw opacity=0] (10.72,-5.15) -- (0,0) -- (10.72,5.15) -- (7.12,0) -- cycle    ;

\draw    (464,46) .. controls (474.67,62.49) and (492.87,69.57) .. (511.29,46.25) ;
\draw [shift={(513,44)}, rotate = 126.16] [fill={rgb, 255:red, 0; green, 0; blue, 0 }  ][line width=0.08]  [draw opacity=0] (10.72,-5.15) -- (0,0) -- (10.72,5.15) -- (7.12,0) -- cycle    ;

\draw (209,143.4) node [anchor=north west][inner sep=0.75pt]    {$b$};
\draw (343,149.4) node [anchor=north west][inner sep=0.75pt]    {$h( z,t)$};
\draw (342,423.4) node [anchor=north west][inner sep=0.75pt]    {$g$};
\draw (320,80.4) node [anchor=north west][inner sep=0.75pt]    {$r$};
\draw (184,450.4) node [anchor=north west][inner sep=0.75pt]    {$z$};
\draw (485,450.4) node [anchor=north west][inner sep=0.75pt]    {$z$};
\draw (457,137.4) node [anchor=north west][inner sep=0.75pt]    {$b$};
\draw (606,80.4) node [anchor=north west][inner sep=0.75pt]    {$r$};
\draw (224,40.4) node [anchor=north west][inner sep=0.75pt]    {$\Omega $};
\draw (520,43.4) node [anchor=north west][inner sep=0.75pt]    {$\Omega $};
\draw (163,244) node [anchor=north west][inner sep=0.75pt]   [align=left] {Outer};
\draw (470,257) node [anchor=north west][inner sep=0.75pt]   [align=left] {Inner};
\end{tikzpicture}
\caption{Model diagram for liquid films flowing along the outer (left)  and inner (right)  surfaces of a rotating cylinder}
\label{fig1}
\end{figure}
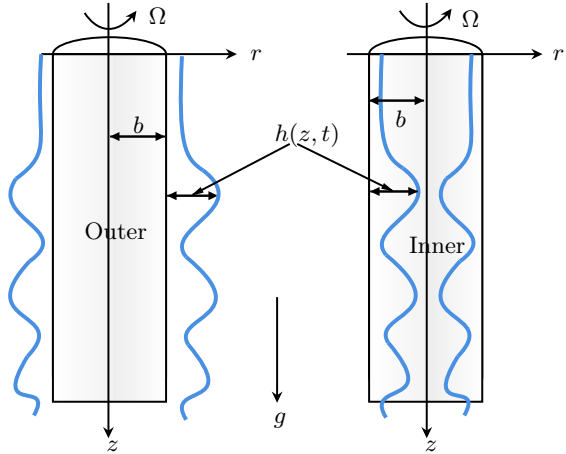

We consider a Newtonian liquid with density $\rho$, dynamic viscosity $\mu$, and surface tension $\sigma$, flowing down a slippery vertical cylinder with radius $r = b$ under gravity $g$. The vertical (axial) coordinate is $z$ (positive downward). We have focused on two scenarios: one where the liquid film flows along the outer surface of the slippery cylinder and the other where it flows along the inner surface. Additionally, we consider the cylinder is rotating about its axis (the $z$-axis) at an angular speed $\Omega$. These systems are displayed in FIG. \ref{fig1}. The thickness of the liquid film at any instant is denoted by $h=h(z,t)$. For liquid flowing along the outer surface of the cylinder, the liquid-air interface is given by $r=b+h(z,t)$, and for liquid flowing along the inner surface of the cylinder, the liquid-gas interface is given by $r=b-h(z,t)$. Thus, we can generally express the equation of the free surface as $r=S(z,t)=b+mh(z,t)$, where $m=\pm1$. The positive sign indicates liquid flow along the outer surface of the cylinder, while the negative sign indicates flow along the inner surface.

\par We consider the two-dimensional hydrodynamic problem, and the governing equations are
\begin{equation}\label{eq1}
r^{-1}\left(ru\right)_r+w_z=0,
\end{equation}
\begin{equation}\label{eq2}
\rho\left(u_t+uu_r+wu_z-r^{-1}v^2\right)=-p_r+\mu\left[r^{-1}\left(ru_r\right)_r+u_{zz}-r^{-2}u\right],
\end{equation}
\begin{equation}\label{eq3}
\rho\left(w_t+uw_r+ww_z\right)=-p_z+\mu\left[r^{-1}\left(rw_r\right)_r+w_{zz}\right]+\rho g.
\end{equation}
Here $p$ represents pressure, and $\mathcal U=(u,v,w)$ represents the radial, azimuthal, and axial fluid velocity components. We assume the azimuthal velocity $v$ remains constant $v=b\Omega$ \cite{souradip2020,chen2004}.

\par At the surface of the cylinder, i.e., at $r=b$, we apply the Navier-slip and no-penetration boundary conditions as
\begin{equation}\label{eq5}
u=0, \quad w=m\widehat\delta w_r,
\end{equation}
where $\widehat\delta$ represents the dimensional slip length.

\par The boundary conditions at the free surface $r=S(z,t)$ are the balance of stresses (tangential and normal) and the kinematic condition, which are given below
\begin{equation}\label{eq7}
\left(u_z+w_r\right)\left(1-S_z^2\right)+2\left(u_r-w_z\right)S_z=0,
\end{equation}
\begin{equation}\label{eq8}
p_a-p+2\mu\left[u_r-\left(u_z+w_r\right)S_z+w_zS_z^2\right]\left(1+S_z^2\right)^{-1}=\sigma\mathcal K,
\end{equation}
\begin{equation}\label{eq9}
u=S_t+wS_z,
\end{equation}
where $\mathcal K=m\left[S_{zz}\left(1+S_z^2\right)^{-1}-r^{-1}\right]\left(1+S_z^2\right)^{-1/2}$ is the curvature and $p_a$ is the atmospheric pressure. 

\par 
We use the following scaling to derive the dimensionless governing equations and boundary conditions
\begin{equation}\label{eq11_scale}
z=\mathcal Lz^*,~(S,r,b)=\mathcal H(S^*,r^*,b^*),~ t=\left(\mathcal L/\mathcal V\right)t^*,~ (u,w)=\mathcal V\left(\epsilon u^*,w^*\right),~ p=p_a+\mathcal P p^*,
\end{equation}
where $\mathcal H$ is the mean thickness of the liquid film, $\mathcal V$ is the velocity scale, $\mathcal L$ is the characteristic length in the axial direction and $\mathcal P$ is the pressure scale and `$*$'  denotes dimensionless variables. We set $\mathcal P=\rho g\mathcal L$, $\mathcal V=\rho g\mathcal H^2/\mu$, and the aspect ratio $\epsilon=\mathcal {H/L}\ll 1$ is a small parameter.

\par We use (\ref{eq11_scale}) in (\ref{eq2})-(\ref{eq9}) with $\mathcal L=\sigma/\left(\rho g\mathcal H\right)$, and obtain the following leading order dimensionless system of equations:
\begin{equation}\label{eq10}
p_r=r^{-1}b^2Ro,~r^{-1}(rw_r)_r=p_z-1,
\end{equation}
\begin{equation}\label{eq11}
u=0,\quad w=m\delta w_r \quad \text{at}\quad r=b,
\end{equation}
\begin{equation}\label{eq12}
w_r=0,\quad p=m\left(r^{-1}-\eta S_{zz}\right),\quad u=S_t+wS_z \quad \text{at}\quad r=S,
\end{equation}
where $Ro=\Omega^2\mathcal H^2/\left(g\mathcal L\right)$ is the rotation number,  $\delta=\widehat\delta/\mathcal H$ is the dimensionless slip parameter and $\eta=\epsilon^2$. To derive (\ref{eq10})-(\ref{eq12}), we have considered $Ro$ and $\delta$ are of order unity. 

\par We solve the system of equations (\ref{eq10})-(\ref{eq12}) and obtain the leading order solutions as
\begin{equation}\label{eq13a}
w=\left(1-p_z\right)\left[\frac{S^2}{2}\ln\left(\frac{r}{b}\right)+\frac{b^2-r^2}{4}+\frac{m\delta}{2b}\left(S^2-b^2\right)\right],
\quad p=m\left(\frac{1}{S}-\eta S_{zz}\right)-b^2Ro\ln\left(\frac{S}{r}\right).
\end{equation}
When the liquid flows along the outer surface of the slippery fiber $(m=1)$ in the absence of rotation $(Ro=0)$, then (\ref{eq13a}) agrees well with Ji et al. \cite{ji2019}.

\par Defining the flow rate $q$ as $q(S)=\int_b^Srwdr$, we obtain the flow rate as given below
\begin{equation}\label{eq13b_flux}
q=\frac{\left(1-p_z\right)}{16}\left[4S^4\ln\left(\frac{S}{b}\right)-3S^4+4S^2b^2-b^4+\frac{4m\delta}{b}\left(S^2-b^2\right)^2\right].
\end{equation}
Equation (\ref{eq13b_flux}) demonstrates that the influence of rotation affects the flow rate via the pressure solution.
\par Considering $\alpha=1/b$, we obtain the interfacial evolution equation from the mass conservative form of the kinematic boundary condition (\ref{eq9}) as
\begin{subequations}\label{eq13_model}
\begin{equation}\label{eq13aaa}
\left(h+\frac{m\alpha}{2}h^2\right)_t+\left[\mathcal M(h)\left(1-\left[\mathcal Z(h)-\eta h_{zz}\right]_z\right)\right]_z=0,
\end{equation}
where the expressions of $\mathcal Z(h)$ and the mobility function $\mathcal M(h)$ are given below:
\begin{equation}\label{eq13aaa1}
\mathcal Z(h)=\frac{m\alpha}{1+m\alpha h}-\frac{Ro}{\alpha^2}\ln\left(1+m\alpha h\right),
\end{equation}
and 
\begin{equation}\label{eq13b}
\mathcal M(h;\alpha,\delta)=\frac{h^3}{3}\phi(m\alpha h)+\frac{h^2}{4}(2+m\alpha h)^2\delta.
\end{equation}
In (\ref{eq13b}), $\phi$ is the shape factor and has the following form
\begin{equation}\label{eq13c}
\phi(Y)=\frac{3}{16Y^3}\left(\left[4\ln(1+Y)-3\right](1+Y)^4+4(1+Y)^2-1\right).
\end{equation}
\end{subequations}
When $Y\rightarrow 0$, we have $\phi(Y)=1+Y+(3/20)Y^2+O\left(Y^3\right)$.

\par Equation (\ref{eq13_model}) represents a fourth-order nonlinear partial differential equation (PDE) governing the evolution of the thickness $h(z,t)$. This model incorporates the influences of surface tension, gravity, and azimuthal instabilities while disregarding the effects of inertia and streamwise viscous dissipation \cite{ruyerquil2009}.  When the cylinder is fixed $(Ro=0)$, and
the thin film flows over the outer surface of a non-slippery $(m=1, \delta=0)$ vertical cylinder, (\ref{eq13_model}) coincides with that obtained by provided by Craster \& Matar \cite{craster2006}. It is important to note that various forms of $\mathcal Z(h)$ are used in the literature to represent the azimuthal curvature of the film \cite{yu2013,craster2006,cm2008}. For a more comprehensive discussion on the different forms of $\mathcal Z(h)$, we refer to the study by Ji et al. \cite{ji2019}.

\par If we rescale the time scale as $t\rightarrow t/\phi(m\alpha)$, then the mobility function (\ref{eq13b}) takes the form
\begin{equation}\label{eq13b_ji}
\mathcal M(h;\alpha,\delta)=\frac{h^3\phi(m\alpha h)}{3\phi(m\alpha)}+\frac{h^2(2+m\alpha h)^2}{4\phi(m\alpha)}\delta.
\end{equation}
In (\ref{eq13b_ji}), setting $m=1$ makes the mobility function $\mathcal M(h)$ exactly match the one found in \cite{ji2019}.

\par In this model, we mainly focus on two parameters: wall slip $(\delta)$ and rotation $(Ro)$. We estimate these parameters' ranges to understand better the significance of the physical mechanisms involved and their interaction. For applications involving flows over hydrophobic surfaces with microscopic texture, the slip length can reach up to the scale of micrometers \cite{rothstein2010}, typically ranging from tens of micrometers to approximately 1 mm \cite{kalliadasisbook}. Consequently, the scaled slip length, $\delta$, would range up to $O\left(10^{-1}\right)$, consistent with values indicated by Chattopadhyay \& Ji \cite{souradip2023aa} and Ding et al. \cite{ding2015}. This order of magnitude for $\delta$ also applies when the substrate is made of a porous material. Here, $\sqrt{\kappa}/B_J$ gives the dimensional slip length, where $\kappa$ is the permeability and $B_J$ is the Beavers-Joseph constant. According to Sadiq et al. \cite{sadiq2010}, with $\sqrt\kappa=8.58\times10^{-5}$ m and $\mathcal H=2.4\times10^{-3}$ m, , and considering the range of $B_J$, the dimensionless slip length ranges from 0.0089 to 0.3575. On the other hand, Ji et al. \cite{ji2019} found the slip length to be in the range of 0.012-0.014, considering the fiber radius $b$ and $\mathcal H$ within $0.1-0.215$ mm and $0.494-0.589$ mm, respectively. However, they considered the slip length from 0 to 1 for theoretical discussions. Similarly, for thin film flows along the outer surface of a porous vertical fiber, Ding \& Liu \cite{ding2011} estimated the dimensionless permeability parameter to be between 0.1 and 0.4. Therefore, we consider the dimensionless slip length $\delta$ to range from 0.1 to 0.4 in the present model.

\par Next, let us determine the range for the second key parameter $Ro$, defined as $Ro=\Omega^2\mathcal H^2/\left(g\mathcal L\right)$. In our model, we define $\mathcal L$, the length scale in the streamwise direction, as $\mathcal L=\mathcal H/\epsilon=\sigma/\left(\rho g\mathcal H\right)$. This implies $\epsilon=\rho g\mathcal H^2/\sigma=Bo$, where $Bo$ represents the Bond number. Typically, $Bo$ is reported to be small in experiments, approximately 0.3 according to Craster \& Matar \cite{craster2006} for a liquid film coating along a vertical fiber. Experimental data from Tan et al. \cite{tan1990} for silicon oil indicates $\rho=10^3$ kg m$^{-3}$ and $\sigma=20.8 \times 10^{-3}$ kg s$^{-2}$. We take $\mathcal H=10^{-4}$ m and $\Omega=2\pi f$ (with $f$ denoting the frequency in Hertz). With $\epsilon=0.3$ and using these values, the value of $Ro$ is approximately $Ro\approx0.0001207291f^2$. Allowing $f$ to vary from 3 to 30, $Ro$ will be approximately 0.001, 0.003, 0.011, 0.027, and 0.108 for $f=3,5,10,15,30$, respectively. Therefore, for our present model, we will consider $Ro$ to be in the range of 0.001 to 0.14 for theoretical discussion. 

\section{Linear stability analysis}\label{sec:3}
In this section, we will perform the linear stability analysis of the problem to get a first
insight into the underlying dynamics. We perturb the PDE (\ref{eq13_model}) as follows: 
\begin{equation}\label{eq13ac_n}
h=1+\widehat h=1+\zeta\exp\left[ikz+\Lambda t\right],
\end{equation}
where $k$ is the wavenumber, $\Lambda\in\mathbb C$ is the growth rate of the perturbation, $i=\sqrt{-1}$ and $\zeta\ll1$ is the disturbance amplitude.
\par We substitute (\ref{eq13ac_n}) in (\ref{eq13_model}) and  ignore the higher-order terms in $\zeta$. Consequently, we obtain the dispersion relation as follows
\begin{subequations}\label{eqq13ac_disp}
\begin{equation}\label{eq13ac_disp}
\Lambda=-c_k\left(\alpha,\delta\right)ik+\frac{k^2}{3\left(1+m\alpha\right)}\left(\frac{\alpha^2}{\left(1+m\alpha\right)^2}+\frac{mRo}{\alpha\left(1+m\alpha\right)}-\eta k^2\right)\chi(\alpha,\delta),
\end{equation}
\begin{equation}\label{eq13ac_disp1n}
c_k\left(\alpha,\delta\right)=\frac{1}{\left(1+m\alpha\right)}\left(\phi(m\alpha)+\frac{m\alpha\phi'(m\alpha)}{3}\right)+(2+m\alpha)\delta,\quad \chi(\alpha,\delta)=\phi(m\alpha)+\frac{3}{4}(2+m\alpha)^2\delta.
\end{equation}
\end{subequations}
While linearizing (\ref{eq13_model}), if we use the mobility function $\mathcal M(h)$ given by (\ref{eq13b_ji}) instead of (\ref{eq13b}), then the expressions of $c_k\left(\alpha,\delta\right)$ and $\chi(\alpha,\delta)$ in (\ref{eq13ac_disp1n}) will take the following form
\begin{equation}\label{eq13ac_disp1n_ji}
c_k\left(\alpha,\delta\right)=\frac{1}{\left(1+m\alpha\right)}\left(1+\frac{m\alpha\phi'(m\alpha)}{3\phi(m\alpha)}\right)+\frac{\left(2+m\alpha\right)}{\phi(m\alpha)}\delta,\quad \chi(\alpha,\delta)=1+\frac{3\left(2+m\alpha\right)^2}{4\phi(m\alpha)}\delta.
\end{equation}
When the cylinder is stationary $(Ro=0)$ and the flow is on the outer side of cylinder $(m=1)$, the dispersion relation (\ref{eq13ac_disp}) with (\ref{eq13ac_disp1n_ji}) agrees well with the dispersion relation obtained by Ji et al. \cite{ji2019}. We note that to derive the dispersion relation (\ref{eq13ac_disp}), we have considered the uniform film solution $h=1$ in (\ref{eq13ac_n}). One can also repeat the above analysis for a uniform thin layer of arbitrary thickness $h=h_0$ \cite{ji2019}.

\par We multiply (\ref{eqq13ac_disp}) by $i$ and use the scaling $i\Lambda\rightarrow c_k\left(c_k/\mathcal Y\right)^{1/3}\widetilde\Lambda$ and $k\rightarrow \left(c_k/\mathcal Y\right)^{1/3}\widetilde k$, where  $\mathcal Y=\eta\chi(\alpha,\delta)/\left(1+m\alpha\right)$. Consequently, (\ref{eqq13ac_disp}) is transformed to the standard form \cite{frenkel1992}
\begin{subequations}
\begin{equation}\label{eq13ac_disp1n_ji_11}
\widetilde\Lambda=\widetilde k+\frac{i\widetilde k^2}{3}\left(\widehat\beta-\widetilde k^2\right),
\end{equation}
\begin{equation}\label{eq13ac_disp1n_ji_111}
\widehat\beta=\frac{\alpha}{\eta\left(1+m\alpha\right)}\left(\frac{\alpha}{\left(1+m\alpha\right)}+\frac{mRo}{\alpha^2 }\right)\left(\frac{c_k}{\mathcal Y}\right)^{-2/3}.
\end{equation}
According to Duprat et al. \cite{duprat2007}, the system's instability becomes absolute when $\widehat\beta>\widehat\beta_{ca}\equiv \left[(9/4)\times\left(-17+7\sqrt{7}\right)\right]^{1/3}\approx1.507$. Therefore, we obtain the following condition on the rotation number $Ro$ 
\begin{equation}\label{eq13ac_disp1n_ji_111a}
mRo>\alpha\left[\eta\left(1+m\alpha\right)\widehat\beta_{ca}\left(\frac{c_k}{\mathcal Y}\right)^{2/3}-\frac{\alpha^2}{\left(1+m\alpha\right)}\right],
\end{equation}
above which the system's instability is absolute.

\par The real part of $\Lambda$, $\Lambda_r = \text{Re}(\Lambda)$, describes the effective growth rate for the system. For $\Lambda_r<0$, the amplitude of the linear growth rate of perturbation decreases, and the system is stable. On the other hand, for $\Lambda_r>0$, the amplitude of the linear growth rate of perturbation increases, and the system is unstable. The expression of $\Lambda_r$ for the present study is given by
\begin{equation}\label{eq13ac_growth}
\Lambda_r=\frac{k^2}{3\left(1+m\alpha\right)}\left[\frac{\alpha^2}{\left(1+m\alpha\right)^2}+\frac{mRo}{\alpha\left(1+m\alpha\right)}-\eta k^2\right]\chi(\alpha,\delta),
\end{equation}
where $\chi(\alpha,\delta)$ is given by (\ref{eq13ac_disp1n_ji}). Equation (\ref{eq13ac_growth}) shows that the effective growth rate $\Lambda_r$ is influenced by both wall slip $\delta$ and rotation number $Ro$.
\begin{figure}[h!]
\centering
\subfloat[]{
\includegraphics[width=0.31\textwidth]{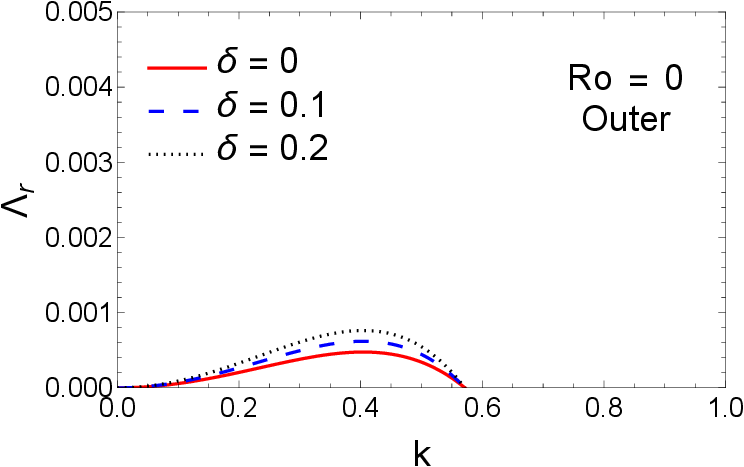} } 
\subfloat[]{
\includegraphics[width=0.31\textwidth]{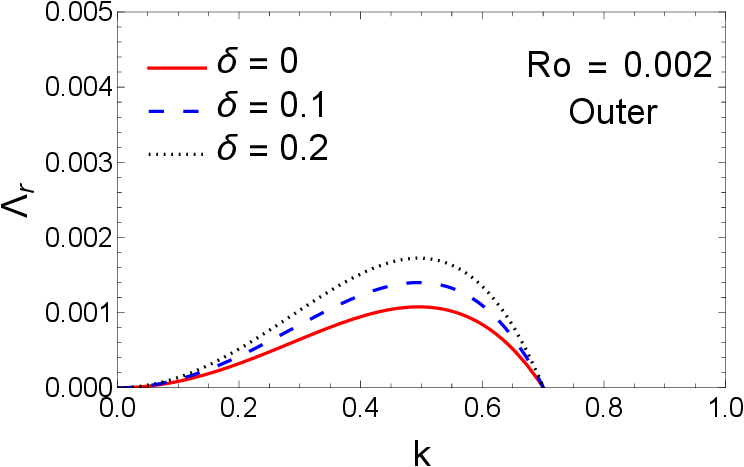} }
\subfloat[]{
\includegraphics[width=0.31\textwidth]{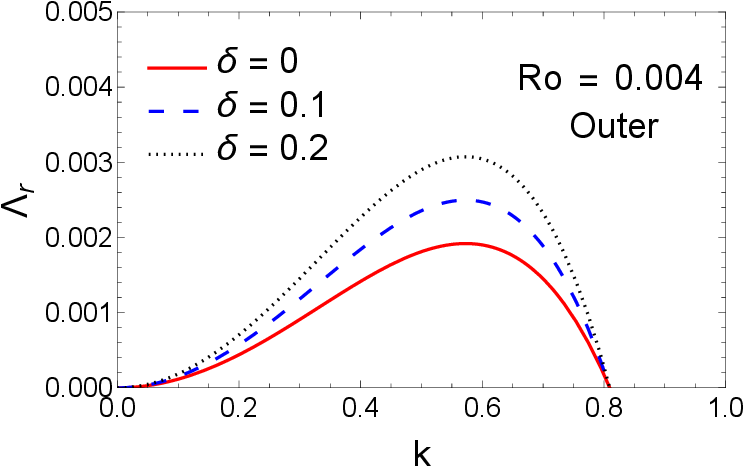} }
\quad
\subfloat[]{
\includegraphics[width=0.31\textwidth]{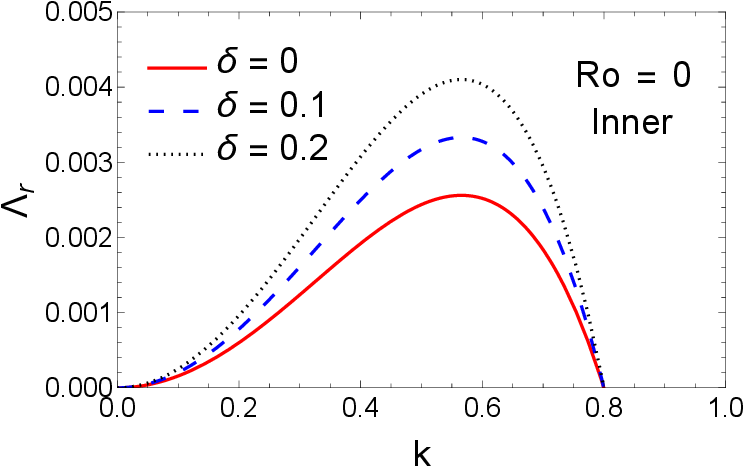} } 
\subfloat[]{
\includegraphics[width=0.31\textwidth]{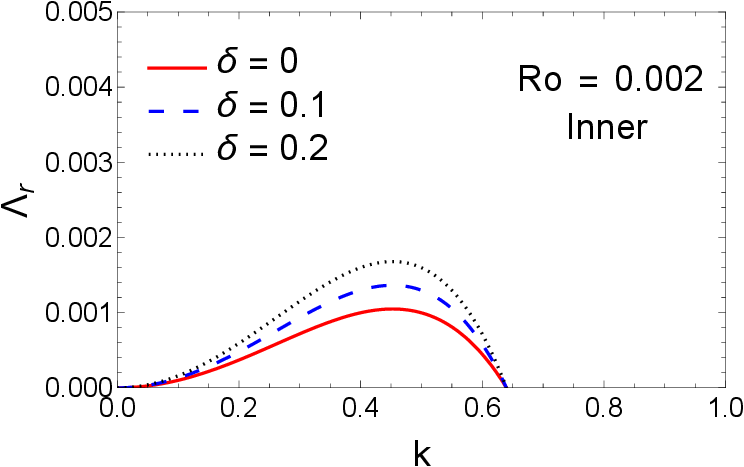} }
\subfloat[]{
\includegraphics[width=0.31\textwidth]{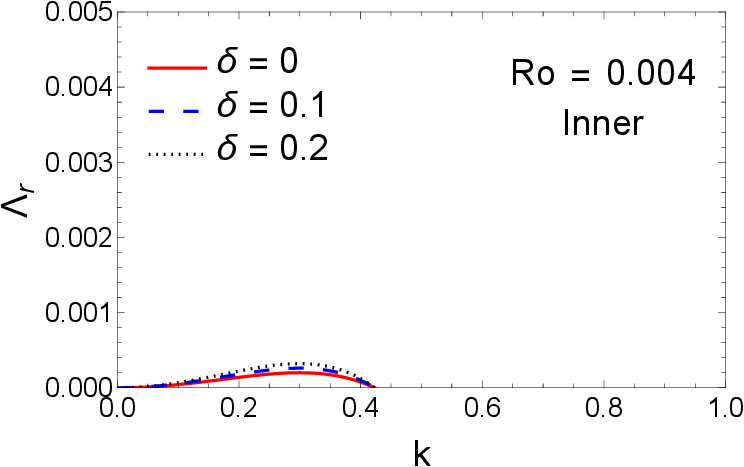} }
\caption{Variation of $\Lambda_r$ with $k$ under different $\delta$ and $Ro$, with $\alpha=1/6$, $\eta=0.04$. Top panel: Flow along outer surface ($m=1$), Bottom: Flow along inner surface ($m=-1$)}	
\label{fig2}
\end{figure}
\par In FIG. \ref{fig2}, we present the linear growth rate curves as functions of the wavenumber $k$. The top panel shows the flow along the outer surface of the cylinder, while the bottom panel illustrates the flow along the inner surface. To analyze the effect of wall slippage, we select three typical $\delta$ values: $\delta = 0, 0.1$, and $0.2$. Our observations indicate that wall slippage promotes flow instability for the cylinder's outer and inner surfaces. This destabilizing influence of wall slippage aligns with findings from previous studies by Ji et al. \cite{ji2019}, Chao et al. \cite{chao2018}, Chattopadhyay \cite{souradip2023aaaa}, and Schwitzerlett et al. \cite{schwitzerlett2023}. Further, FIGs. \ref{fig2}b and \ref{fig2}c illustrate the impact of rotation on the flow along the outer surface of the cylinder. To explore the effects of rotation, we select two typical $Ro$ values: $Ro = 0.002$ and $0.004$. By comparing FIGs. \ref{fig2}a to \ref{fig2}c, we observe that as $Ro$ increases, flow instability also increases, indicating that rotation destabilizes the flow along the outer surface of the cylinder. This finding aligns with the results presented by Liu \& Ding \cite{liu2020} and Mukhopadhyay et al. \cite{souradip2020}. In contrast, FIGs. \ref{fig2}e and \ref{fig2}f examine the effect of rotation on the flow along the inner surface of the cylinder, with all parameters matching those in the top panel of FIG. \ref{fig2}. Comparing FIGs. \ref{fig2}d to \ref{fig2}f reveals that increasing $Ro$ decreases flow instability, demonstrating that rotation stabilizes the flow along the inner surface of the cylinder. Thus, our linear stability analysis concludes that rotation exerts completely different influences depending on whether the film flows along the outer or inner surface of the cylinder/fiber. Physically, when the film flows along the outer surface, the centrifugal force destabilizes the flow by pushing fluid away from the cylinder wall. Conversely, when the film flows along the inner surface, the centrifugal force stabilizes the flow by drawing fluid toward the cylinder wall. Moreover, the linear growth rate profiles show that for given $\delta$ and $Ro$, a critical wavenumber exists below which instability increases, after which it decreases.

\par From (\ref{eq13ac_growth}), we observe that if 
\begin{equation}\label{eq13ac_growth1}
\frac{\alpha^3+mRo\left(1+m\alpha\right)}{\alpha\eta\left(1+m\alpha\right)^2}\leq\eta k^2,
\end{equation}
the long-wave instability can be completely impeded by rotation. Consequently, we can define a sufficient condition for the system to be stable in the long-wave range as
\begin{equation}\label{eq13ac_growth2}
mRo\leq -\frac{\alpha^3}{\left(1+m\alpha\right)}.
\end{equation}
We further obtain the cut-off wavenumber $\left(k=k_c\right)$ and the most unstable
mode $\left(k=k_m\right)$ as
\begin{equation}\label{eq13ac_growth3}
k_c=\sqrt{\frac{\alpha^3+mRo\left(1+m\alpha\right)}{\alpha\eta\left(1+m\alpha\right)^2}},\quad k_m=\frac{k_c}{\sqrt{2}}.
\end{equation}
Equation (\ref{eq13ac_growth3}) demonstrates that the slip length $\delta$ does not influence the critical wavenumber $k_c$ for both outer and inner surface flow of the vertical cylinder, and consequently $k_m$. This observation aligns with the findings of Chao et al. \cite{chao2018}, who studied the flow along the outer surface of a slippery cylinder under non-isothermal conditions. They reported that the critical wavenumber is independent of the slip length without thermal effects. 
\end{subequations}

\begin{figure}[h!]
\centering
\subfloat[]{
\includegraphics[width=0.31\textwidth]{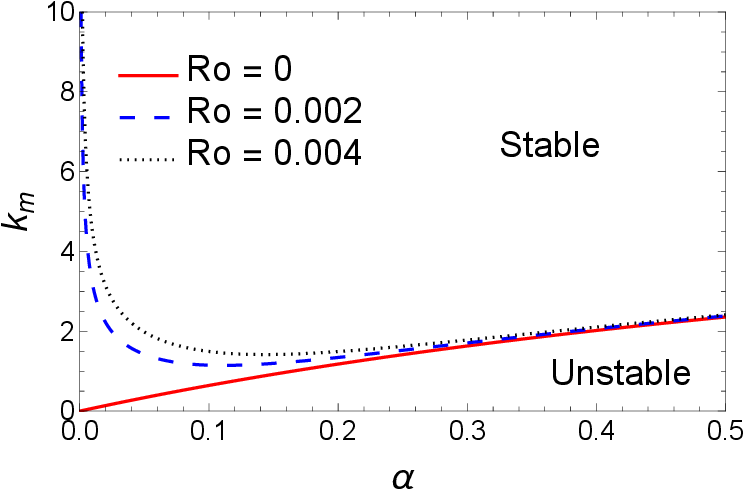} } 
\subfloat[]{
\includegraphics[width=0.31\textwidth]{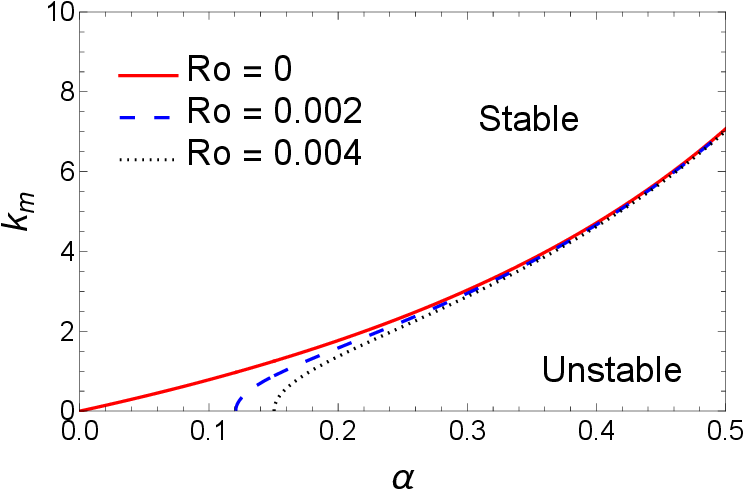} }
\caption{Influence of $Ro$ on the most unstable mode with $\eta=0.01$. (a) Flow along outer surface ($m=1$) and (b) Flow along inner surface ($m=-1$)}	
\label{fig3}
\end{figure}
\par In FIG. \ref{fig3}, we illustrate how the fastest growing mode $k = k_m$ is influenced by rotation ($Ro$) as the cylinder radius $(b = 1/\alpha)$ changes. FIG. \ref{fig3}a depicts the flow along the outer surface of the cylinder, while FIG. \ref{fig3}b shows the flow along the inner surface. The $Ro$ values are consistent with those used in FIG. \ref{fig2}. FIG. \ref{fig3}a demonstrates that as $Ro$ increases, the unstable zone expands, whereas FIG. \ref{fig3}b shows that as $Ro$ increases, the unstable zone contracts. These findings confirm the dual role of rotation $Ro$, depending on whether the film flows along the outer or inner surface of the cylinder. Additionally, both figures reveal that the unstable region consistently enlarges with increasing $\alpha$. In other words, flow instability is enhanced as $\alpha$ increases (i.e., the radius of the cylinder decreases). This observation aligns with previous studies by Craster \& Matar \cite{craster2006} and Mukhopadhyay et al. \cite{souradip2020}. This phenomenon occurs regardless of the presence of rotation.

\par The imaginary part of $\Lambda$, $\Lambda_i = \text{Im}(\Lambda)$, is $\Lambda_i=-kc_k\left(\alpha,\delta\right)$, where $c_k\left(\alpha,\delta\right)$ is given by (\ref{eq13ac_disp1n_ji}). The rotation does not influence $\Lambda_i$ but is affected by the wall slippage $\delta$ for outer and inner flow cases. We further obtain the linear wave speed $c_{\text{lin}}$ as $c_{\text{lin}}=-\Lambda_i/k=c_k\left(\alpha,\delta\right)$. Therefore, the linear wave speed depends only on the wall slippage, not the rotation. However, whether the rotation influences the nonlinear wave speed for outer and inner surface flows remains unknown. 
we will discuss the impact of rotation on nonlinear wave speeds in Section \ref{sec:5}. 

\section{Weakly nonlinear stability analysis}\label{sec:4}
The linear study fails to capture the accurate flow behavior when the perturbed waves grow to a finite amplitude. Therefore, a weakly nonlinear stability analysis is necessary to determine the behavior of the nonlinear wave near criticality. In this section, we focus on understanding the influence of wall slippage and rotation on the weakly nonlinear stability of the model (\ref{eq13_model}), following \cite{souradip2022prf, souradip2023aa, oron2004, souradipjem, samanta2008, Desai2023b,souradippof2021a}. We express the PDE (\ref{eq13_model}) as:
\begin{subequations}\label{eq13_model_weakly}
\begin{equation}
h_t+\gamma_1(h)h_z+\gamma_2(h)h_{zz}+\gamma_3(h)h_{zzzz}+\gamma_4(h)h_z^2+\gamma_5(h)h_zh_{zzz}=0,
\end{equation}
where the coefficients $\gamma_i,i=1,\ldots,5$ are as follows:
\begin{equation}
\gamma_1(h;\alpha,\delta)=\frac{\alpha\Theta}{16m(1+m\alpha h)},
\quad \gamma_2(h;\alpha,\delta,Ro)=\frac{\alpha^2\theta}{(1+m\alpha h)^2}\left(\frac{Ro}{16\alpha^2}+\frac{\gamma_1}{\Theta}\right), \quad \gamma_3(h;\alpha,\delta)= \frac{\gamma_1\eta\theta}{\Theta},
\end{equation}
\begin{equation}
\gamma_4(h;\alpha,\delta,Ro)=\frac{\alpha^2\left[\Theta+m\alpha\left(h\Theta-\theta\right)\right]}{(1+m\alpha h)^3}\left(\frac{Ro}{16\alpha^2}+\frac{\gamma_1}{\Theta}\right)-\frac{\alpha^4\theta}{16(1+m\alpha h)^4},\quad \gamma_5(h;\alpha,\delta)=\eta\gamma_1,
\end{equation}
\begin{equation}
\theta(h;\alpha,\delta)=\frac{m\alpha h\left(2+m\alpha h\right)\left[m\alpha h(2+m\alpha h)\left(4m\alpha\delta-3\right)-2\right]+4\left(1+m\alpha h\right)^4\ln(1+m\alpha h)}{\alpha^4},
\end{equation}
\begin{equation}
\Theta(h;\alpha,\delta)=\frac{8(1+m\alpha h)\left[m\alpha h(2+m\alpha h)(2\alpha\delta-m)+2m(1+m\alpha h)^2\ln(1+m\alpha h)\right]}{\alpha^3}.
\end{equation}
\end{subequations}
To investigate the nonlinear effect of the present model, we use the method of multiple scales as
\begin{equation}\label{eqqta}
\partial_{t}\rightarrow\partial_{t}+\sum_{j=1}\psi^j\partial_{t_j},
\quad
\partial_{z}\rightarrow\partial_{z}+\sum_{j=1}\psi^j\partial_{z_{j}},\quad \widehat h=\sum_{j=1}\psi^j\widehat h_j.
\end{equation}
where $\psi=k-k_c$ is a small perturbation parameter.

\par Inserting (\ref{eq13ac_n}) in (\ref{eq13_model_weakly}), considering terms up to $O\left(\widehat h^3\right)$ and using (\ref{eqqta}) gives the following Landau-type equation for the perturbation amplitude $\zeta$ \cite{souradip2022prf,souradip2023aa,oron2004,souradipjem,samanta2008,Desai2023b}:
\begin{equation}\label{eq73}
\dfrac{\partial\zeta}{\partial t_{2}}+iJ_0\dfrac{\partial\zeta}{\partial z_{1}}+J_{1}\dfrac{\partial^{2}\zeta}{\partial z_{1}^{2}} -\psi^{-2}\Lambda_r\zeta + \left(J_{2}+iJ_{4} \right)|\zeta|^{2}\zeta=0, 
\end{equation}
We refer to the studies by Oron \& Gottlieb \cite{oron2004} and Chattopadhyay et al. \cite{souradip2021b} to get a detailed derivation of (\ref{eq73}) from (\ref{eq13_model_weakly}). The coefficients $J_0$, $J_1$, $J_2$ and $J_4$ are given below:
\begin{subequations}
\begin{equation}
J_0=2k\left(\Gamma_1-2\Gamma_3k^2\right)\psi^{-1}, \quad J_1=\Gamma_1-6\Gamma_3k^2,
\quad 
J_2=k^2\left(\mathcal E\mathcal V-\frac{3}{2}\Gamma_2''+\frac{3}{2}\Gamma_3''k^2+\Gamma_4'-\Gamma_5'k^2\right)-\Gamma_1'\mathcal F,
\end{equation}
\begin{equation}
\mathcal V=-5\Gamma_2'+17\Gamma_3'k^2+4\Gamma_4-10\Gamma_5k^2,\quad J_4=\Gamma_1'k\mathcal E+\frac{\Gamma_1''}{2}k+\mathcal Vk^2\mathcal F,
\end{equation}
\begin{equation}
\mathcal E=\frac{\Gamma_2'-\Gamma_3'k^2+\Gamma_4-\Gamma_5k^2}{4\left(4\Gamma_3k^2-\Gamma_2\right)},\quad \mathcal F=\frac{-\Gamma_1'}{4\left(4\Gamma_3k^2-\Gamma_2\right)}.
\end{equation}
Here, the $\Gamma_i$ and its derivatives are evaluated from $\gamma_i$ at $h=1$.
\end{subequations}
\par When there is no spatial modulation (filtered wave), the equation (\ref{eq73}) yields
\begin{equation}\label{eq73aa}
\dfrac{\partial\zeta}{\partial t_{2}} -\psi^{-2}\Lambda_r\zeta + \left(J_{2}+iJ_{4} \right)|\zeta|^{2}\zeta=0. 
\end{equation}
In this case, the solution for the perturbation amplitude $\zeta$ can be expressed in a simple form as \cite{oron2004}: $\zeta=a\left(t_2\right)\exp\left[-ib\left(t_2\right)t_2\right]$. We use this solution of $\zeta$ in (\ref{eq73aa}). Separating the real and imaginary parts, we obtain the following
\begin{equation}\label{eq73aaa}
\frac{\partial a}{\partial t_2}=a\left(\frac{\Lambda_r}{\psi^2}-J_2a^2\right).
\end{equation}
In (\ref{eq73aaa}), the second term on the right-hand side is contributed by system nonlinearities. In the linear stability analysis, the exponential growth of the linear disturbance is decided by the sign of $\Lambda_r$. Beyond the linear regime, the perturbation dynamics depend solely on the sign of $J_2$. A positive $J_2$ indicates a supercritical bifurcation, while a negative $J_2$ indicates a subcritical bifurcation. Therefore, based on the signs of $\Lambda_r$ and $J_2$, we identify four zones: subcritical stable $\left(J_2<0,\Lambda_r<0\right)$, subcritical unstable $\left(J_2<0,\Lambda_r>0\right)$, supercritical stable $\left(J_2>0,\Lambda_r<0\right)$ and supercritical unstable $\left(J_2>0,\Lambda_r>0\right)$ \cite{oron2004,sadiq2010}. These zones are labeled 1, 2, 3, and 4. This discussion highlights that understanding the effects of nonlinear terms on the stability of the flow system requires characterizing various flow states using $J_2$, which can be determined through weakly nonlinear stability analysis.

\par Moreover, when $a$ is independent of $t_2$, we can obtain the amplitude for a saturated wave (threshold amplitude) from (\ref{eq73aaa}) as \cite{souradip2020}
\begin{equation}
\psi a=\sqrt{\frac{\Lambda_r}{J_2}}.
\end{equation}

\begin{figure}[h!]
\centering
\subfloat[]{
\includegraphics[width=0.32\textwidth]{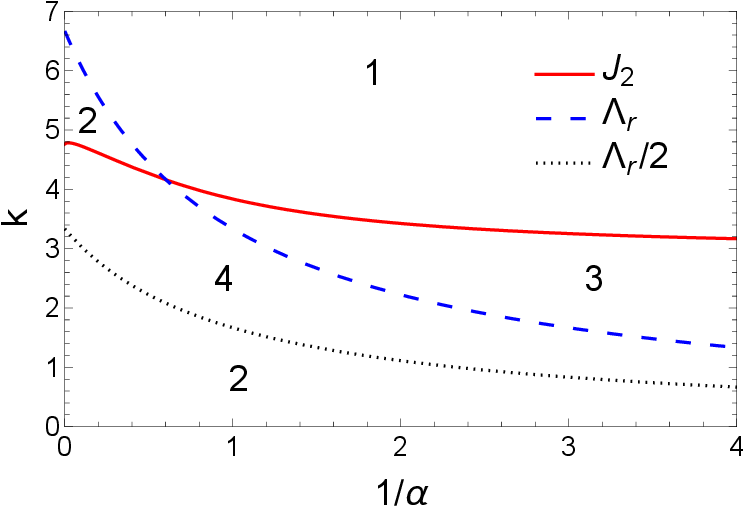} }
\subfloat[]{
\includegraphics[width=0.32\textwidth]{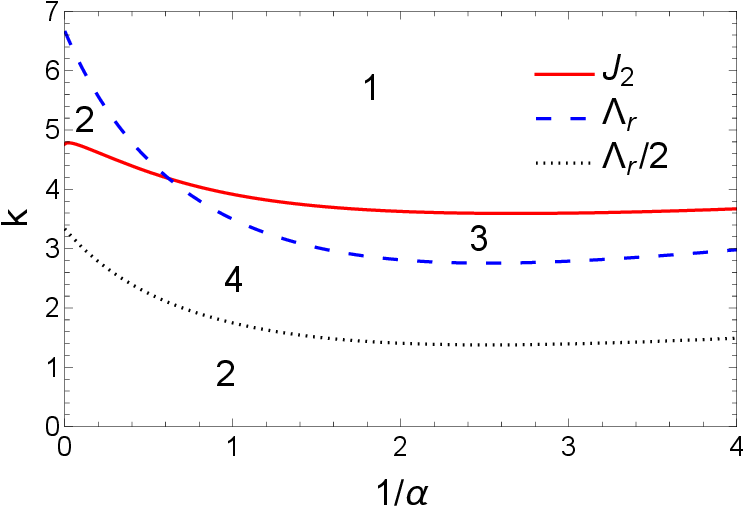} }
\quad
\subfloat[]{
\includegraphics[width=0.32\textwidth]{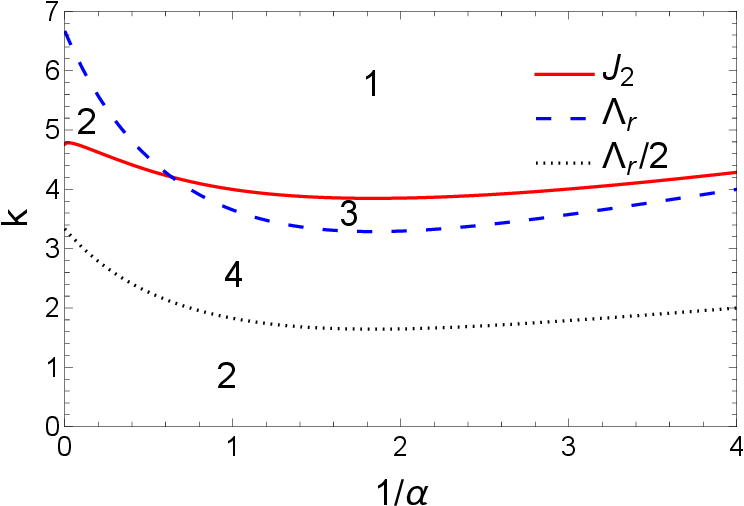} }
\subfloat[]{
\includegraphics[width=0.34\textwidth]{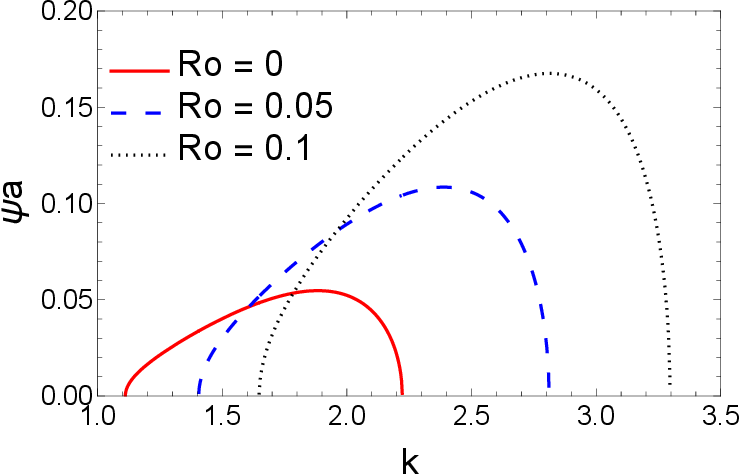} }
\caption{Effect of $Ro$ when the film flows along the outer surface of the cylinder  $(m=1)$ for $\eta = 0.0225$. (a-c) Neutral stability curve for $\left(\delta,Ro\right)=(0.2,0)$ in (a);  $\left(\delta,Ro\right)=(0.2,0.05)$ in (b); $\left(\delta,Ro\right)=(0.2,0.1)$ in (c). (d) Threshold amplitude in the supercritical unstable region with $\alpha=0.5$ and $\delta=0.2$}	
\label{figwnsa}
\end{figure}
FIG. \ref{figwnsa} is plotted for thin film flows along the outer surface of the cylinder. FIG. \ref{figwnsa}a shows a stationary cylinder $(Ro=0)$ with wall slippage $(\delta=0.4)$.  while FIG. \ref{figwnsa}b shows a rotating cylinder $(Ro=0.02)$ with the same slippage. Rotation expands the subcritical unstable zone (zone 2) and contracts the supercritical stable zone (zone 3). In FIG. \ref{figwnsa}c, increasing the rotation to $Ro=0.04$ further amplifies zone 2 and reduces zone 3, confirming that rotation destabilizes the film flow, with instability increasing with wall slippage. In FIG. \ref{figwnsa}d, we illustrate the impact of rotation on the threshold amplitude in the supercritical unstable zone $\left(J_2>0,\Lambda_r>0\right)$ with a slippery cylinder wall $(\delta>0)$. We notice that the threshold amplitude rises with increased rotation. Moreover, for a fixed $Ro$, the threshold amplitude initially increases until reaching a critical wavenumber $k=k^\dagger$. Beyond $k^\dagger$, the threshold amplitude decreases.

\begin{figure}[h!]
\centering
\subfloat[]{
\includegraphics[width=0.32\textwidth]{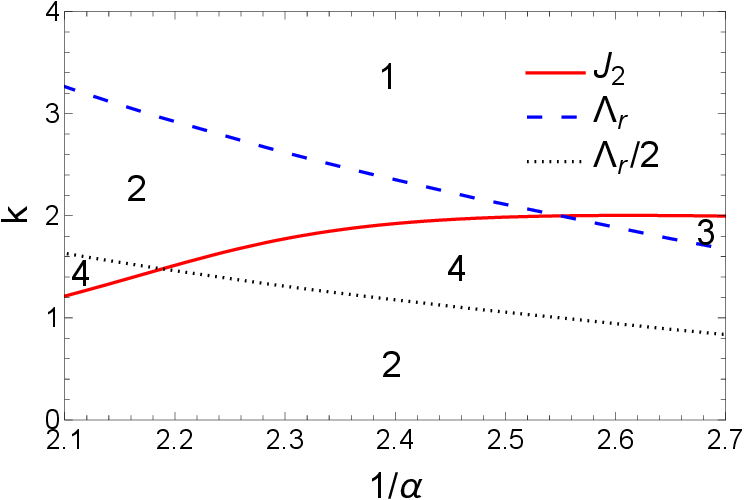} }
\subfloat[]{
\includegraphics[width=0.32\textwidth]{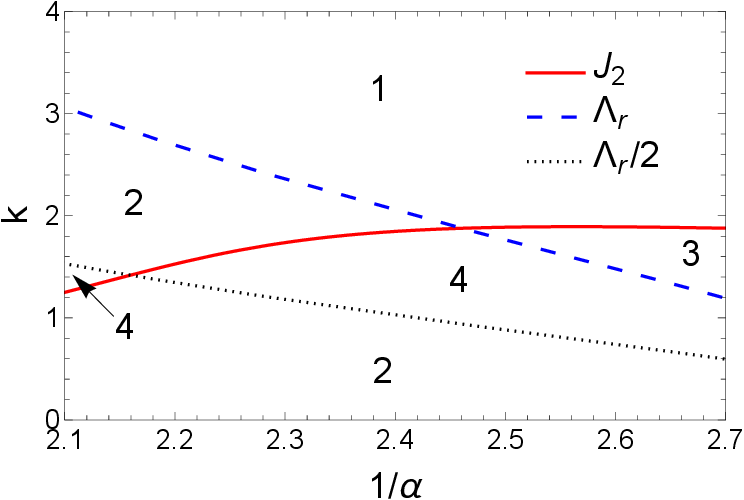} }
\quad
\subfloat[]{
\includegraphics[width=0.32\textwidth]{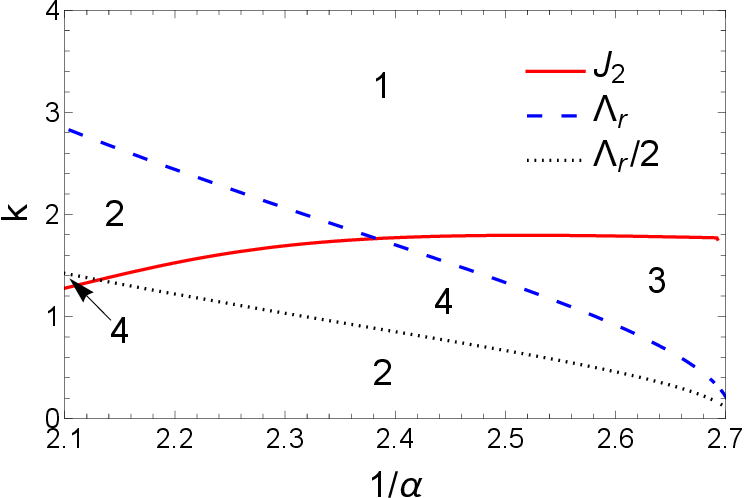} }
\subfloat[]{
\includegraphics[width=0.33\textwidth]{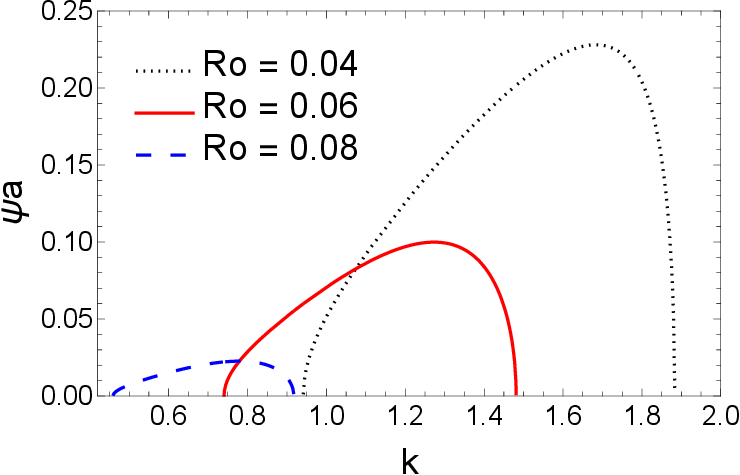} }
\caption{Effect of $Ro$ when the film flows along the inner surface of the cylinder  $(m=-1)$ for $\eta = 0.0625$. (a-c) Neutral stability curve for $\left(\delta,Ro\right)=(0.4,0.04)$ in (a);  $\left(\delta,Ro\right)=(0.4,0.06)$ in (b); $\left(\delta,Ro\right)=(0.4,0.08)$ in (c). (d) Threshold amplitude in the supercritical unstable region with $\alpha=1/2.6$ and $\delta=0.4$}	
\label{figwnsa_n}
\end{figure}
In FIG. \ref{figwnsa_n}, we illustrate the impact of rotation $(Ro)$ on the supercritical and subcritical stability regions when liquid flows along the inner surface of a slippery $(\delta>0)$ cylinder. We set the slip length $\delta$ to 0.4 and select three typical $Ro$ values: $Ro=0.04,0.06$ and 0.08. Comparing FIGs. \ref{figwnsa_n}a to \ref{figwnsa_n}c, we observe that as the $Ro$ value increases, zone 3 (supercritical stable) expands, while zone 2 (subcritical unstable) contracts. In the supercritical stable region $\left(J_2>0,\Lambda_r<0\right)$, , finite-amplitude disturbances remain always stable. Conversely, in the subcritical unstable region $\left(J_2<0,\Lambda_r>0\right)$, both linear and nonlinear instabilities increase, making the system always unstable. FIGs. \ref{figwnsa_n}(a-c) demonstrate that zone 3, where finite-amplitude disturbances are always stable, enlarges with higher $Ro$, whereas zone 2, where instability consistently increases, diminishes with higher $Ro$. Thus, the stabilizing effect of $Ro$ is consistent with the results from linear stability analysis. In FIG. \ref{figwnsa_n}d, we demonstrate how rotation affects the threshold amplitude in the supercritical unstable zone $\left(J_2>0,\Lambda_r>0\right)$ with a slippery cylinder wall $(\delta>0)$. We observe a reduction in the threshold amplitude with increased rotation, contrary to the scenario where the liquid flows along the outer surface of the cylinder (see FIG. \ref{figwnsa}d). Similarly to FIG. \ref{figwnsa}d, we observe the existence of a critical wavenumber $k=k^\dagger$ for a given $Ro$. Below $k^\dagger$, the threshold amplitude increases, whereas above $k^\dagger$, the threshold amplitude decreases.

\section{Traveling wave solutions and their stabilities}\label{sec:5}
In this section, we investigate how rotation affects the characteristics of traveling wave (TW) solutions in the presence of wall slippage on both the outer and inner surfaces of the flow governed by (\ref{eq13_model}). Prior works \cite{ji2022,taranets2024,ji2019,schwitzerlett2023} have studied TW solutions for thin films flowing along either the inner or outer surface of a vertical cylinder, with an emphasis on the impact of the substrate geometry and slip length. The linear stability analysis in Section \ref{sec:3} concludes that the linear wave speed is influenced by wall slippage but not by rotation for outer and inner surface flows. Here, we explore how the slippery length and rotation affect the nonlinear wave speed. 

\subsection{Traveling wave solutions}\label{sec:5a}
We consider the model on a periodic domain $0\leq z\leq L$ and introduce a change of variables to the reference frame of the traveling wave (TW) following \cite{ji2019},
\begin{equation}\label{eqqtw}
\xi=z-ct,~s=t,~ h(z,t)=\widetilde h(\xi,s),
\end{equation}
where $c$ denotes the TW speed. 
Substituting (\ref{eqqtw}) into (\ref{eq13_model}), we obtain the PDE for $\widetilde h(\xi,s)$ in the moving reference frame,
\begin{equation}\label{eq13aaatw}
\left(\widetilde h+\frac{m\alpha}{2}\widetilde h^2\right)_s-c\left(\widetilde h+\frac{m\alpha}{2}\widetilde h^2\right)_\xi+\left[\mathcal M\left(\widetilde h\right)\left(1-\left[\mathcal Z\left(\widetilde h\right)-\eta \widetilde h_{\xi\xi}\right]_\xi\right)\right]_\xi=0.
\end{equation}
The TW solution $H(\xi)$ represents a steady state of the PDE (\ref{eq13aaatw}) and satisfies the fourth-order nonlinear ODE:
\begin{equation}\label{eq13aaatw1}
c\left(H+\frac{m\alpha}{2}H^2\right)_\xi=\left[\mathcal M\left(H\right)\left(1-\left[\mathcal Z\left(H\right)-\eta H_{\xi\xi}\right]_\xi\right)\right]_\xi.
\end{equation}
This forms a nonlinear eigenvalue problem, where the propagation speed $c$ acts as the eigenvalue. We employ centered finite differences and Newton's method to solve the nonlinear ODE \eqref{eq13aaatw1}, treating the speed $c$ as an unknown. For local uniqueness, we specify $H\left(\xi_0\right)=H_0$ for some $0\leq\xi_0\leq L$ and enforce a mass constraint:
\begin{equation}\label{eqq_tw2}
\int_0^L\left(H+\frac{m\alpha}{2}H^2\right) ~d\xi=M_0,
\end{equation}
where $M_0$ is the given mass and $L$ is the domain size. Using the continuation method, we numerically trace a family of TW profiles parametrized by the rotation parameter $Ro$ and the slip length $\delta$, with a prescribed mass $M_0$, domain size $L$, and other system parameters. For simplicity, we focus on cases where a one-period solution fits within the domain.

\begin{figure}[h!]
\centering
\subfloat[$m=-1$, $Ro=0.004$]{
\includegraphics[width=0.32\textwidth]{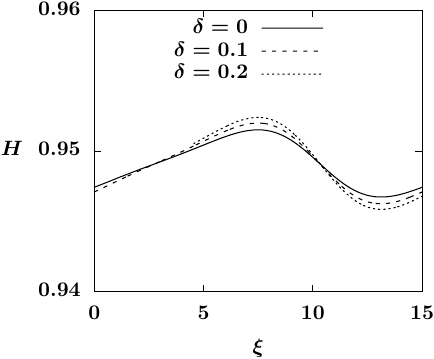} }
\subfloat[$m=-1$, $\delta=0.2$]{
\includegraphics[width=0.32\textwidth]{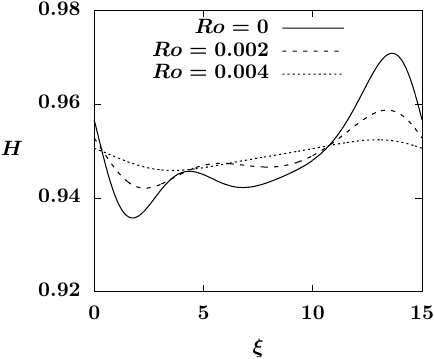} }
\quad
\subfloat[$m=1$, $Ro=0.02$]{
\includegraphics[width=0.33\textwidth]{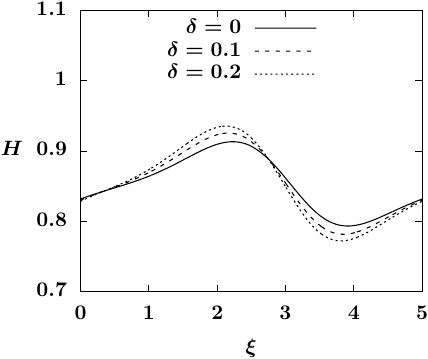} }
\subfloat[$m=1$, $\delta=0$]{
\includegraphics[width=0.32\textwidth]{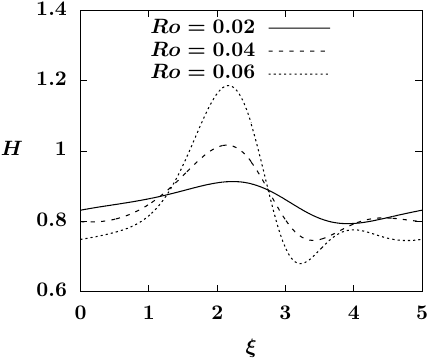} }
\caption{Traveling wave profiles satisfying the fourth-order ODE \eqref{eq13aaatw1} with varying $\delta$ and $Ro$ values. For films flowing along the inner surface of the cylinder ($m=-1$), (a) shows the effect of $\delta$ with a fixed $Ro = 0.004$, and (b) shows the effect of $Ro$ with a fixed $\delta=0.2$. For traveling waves along the outer surface of the cylinder ($m=1$),  (c) compares the TW profiles with a varying $\delta$ and a fixed $Ro = 0.02$, and (d) presents the effects of $Ro$ with a fixed $\delta = 0$. The other system parameters are $\alpha = 1/6$, $\eta = 0.04$. For the case $m = 1$, we set $(M_0, L) = (4.569, 5)$. For the case $m = -1$, we have $(M_0, L) = (13.124, 15)$.}	
\label{TWS_comparison}
\end{figure}

\begin{figure}[h!]
\centering
\subfloat[]{
\includegraphics[width=0.3\textwidth]{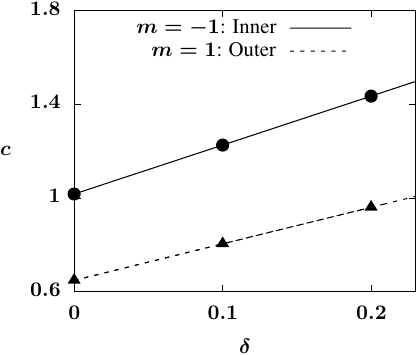} }
\subfloat[]{
\includegraphics[width=0.31\textwidth]{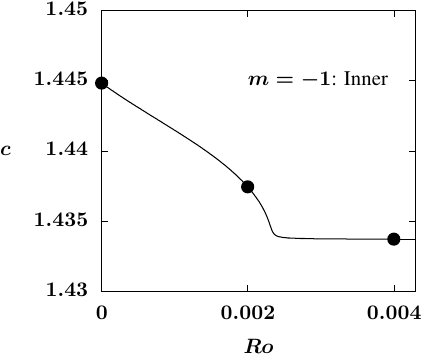} }
\quad
\subfloat[]{
\includegraphics[width=0.31\textwidth]{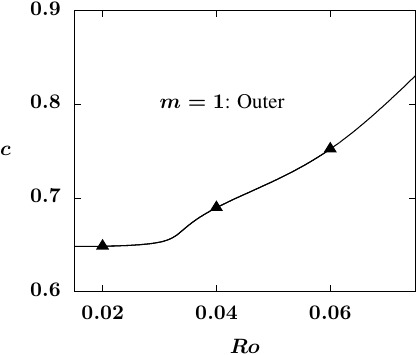} }
\caption{The corresponding propagation speeds of the TW solutions shown in FIG.~\ref{TWS_comparison}. (a) The speed increases as the slip length $\delta$ increases for films flowing along the cylinder's inner and outer surfaces. (b - c) A larger rotation number $Ro$ leads to a decreased (increased) TW speed for the inner (outer) surface flow case. The markers indicate the speeds for TW profiles presented in FIG.~\ref{TWS_comparison} in each case, circles for the $m=-1$ cases, and triangles for the $m=1$ cases.}	
\label{TWS_speed_comparison}
\end{figure}

\par In FIGs.~\ref{TWS_comparison}-\ref{TWS_speed_comparison}, we present numerical solutions for typical TW profiles and their propagation speeds for liquid films flowing along both the inner ($m=-1$) and outer ($m=1$) surfaces of the cylinder. 
For all the inner surface TW profiles ($m=-1$), we set the domain size $L = 15$ and the mass constraint $M_0 = 13.124$. For the outer surface TW profiles ($m=1$), we set the domain size $L = 5$ and the mass constraint $M_0 = 4.569$. FIGs.~\ref{TWS_comparison}a and \ref{TWS_comparison}c show that for both cases with $m=\pm 1$, when other system parameters are fixed, stronger slip effects yield single-peak traveling waves with larger amplitudes. This is consistent with the linear stability analysis results, which show that slip effects can lead to enhanced wave instability. 
The associated speeds of the TW solutions are plotted in FIG.~\ref{TWS_speed_comparison}a, showing that a larger slip length leads to larger wave speeds for both inner and outer surface TWs. 

\par FIGs.~\ref{TWS_comparison}b and \ref{TWS_comparison}d reveal the different roles of rotation in TW solutions for inner and outer surface flows. With other system parameters identical, a larger rotation number leads to reduced variations in the TW profiles for inner surface flows and enlarged amplitudes for outer surface TW profiles. Correspondingly, FIGs.~\ref{TWS_speed_comparison}b and \ref{TWS_speed_comparison}c show that the rotation effects lead to reduced TW speeds for inner surface flows and induce higher speeds for outer surface flow TW solutions.

\subsection{Stability of traveling wave solutions}\label{sec:5b}
Previous studies by Ji et al. \cite{ji2019} and Liu and Ding \cite{liu2020} examined the stability of traveling wave (TW) solutions for a vertical cylinder under various conditions, focusing on liquid film flow along the outer surface. However, when the thin film flows along a rotating cylinder's outer or inner surface in the presence of wall slippage, the TW solutions and their stability remain unexplored. Following \cite{ji2019}, we conduct a stability analysis of TW solutions for the present model in this section, focusing on perturbations with the same period. 

\par Let us consider a positive periodic TW solution $H(\xi)$ over the domain $\xi\in[0,L]$. We perturb it by setting $\widetilde h\left(\xi,s\right)=H(\xi)+\Upsilon\Psi(\xi)e^{\lambda s}$, where $\Upsilon\ll1$ and $\Psi(\xi)$ is also $L$-periodic. We linearize (\ref{eq13aaatw}) around the steady state $H(\xi)$ and obtain the $O(\Upsilon)$ equation 
\begin{equation}\label{eq4.5n}
\lambda\Psi=\mathbb L\Psi,
\end{equation}
where the linear operator $\mathbb L$ in (\ref{eq4.5n}) is obtained as
\begin{equation*}
\mathbb L\Psi\equiv c\left(\Psi_\xi+\frac{m\alpha H_\xi\Psi}{1+m\alpha H}\right)+\frac{1}{1+m\alpha H}\left[\mathcal M(H)\left(\left[\mathcal Z(H)\right]_\xi\Psi-\eta\Psi_{\xi\xi}\right)_\xi\right]_\xi
\end{equation*}
\begin{equation}\label{eq4.6n}
-\frac{1}{1+m\alpha H}\left(\left[\mathcal M(H)\right]_\xi\left[1-\left(\mathcal Z(H)-\eta H_{\xi\xi}\right)_\xi\right]\Psi\right)_\xi.
\end{equation}
In (\ref{eq4.6n}), setting $m=1$ describes the case where the liquid flows over the outer surface of a rotating vertical cylinder with wall slippage. Conversely, setting $m=-1$ corresponds to the liquid flowing along the inner surface of the rotating vertical cylinder with wall slippage. The effect of rotation $(Ro)$ is incorporated through the expression $\mathcal Z(H)$, while the influence of wall slippage $(\delta)$ is introduced via the mobility function $\mathcal M(H)$. When the flow is along the outer surface of the cylinder $(m=1)$ and the cylinder is not rotating $(Ro=0)$, (\ref{eq4.6n}) agrees well with Ji et al. \cite{ji2019}. Here, $\Psi(\xi)$ denotes a normalized eigenfunction associated with an eigenvalue $\lambda$, where $\|\Psi\|_2=1$. If there exists an eigenvalue $\lambda$ with a positive real part in  (\ref{eq4.5n}), then the periodic TW solution $H(\xi)$ is unstable. 

\begin{figure}
\centering
\subfloat[]{
\includegraphics[width=0.3\textwidth]{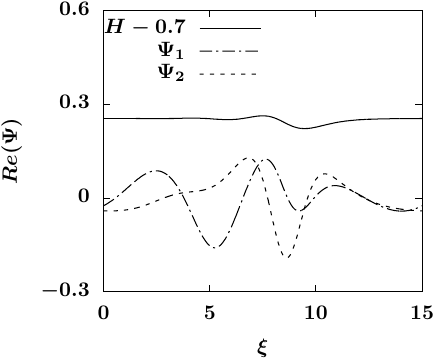} }
\subfloat[]{
\includegraphics[width=0.31\textwidth]{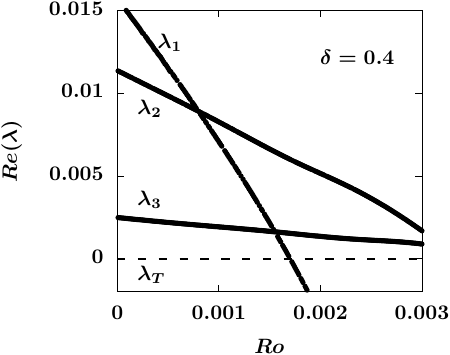} }
\subfloat[]{
\includegraphics[width=0.31\textwidth]{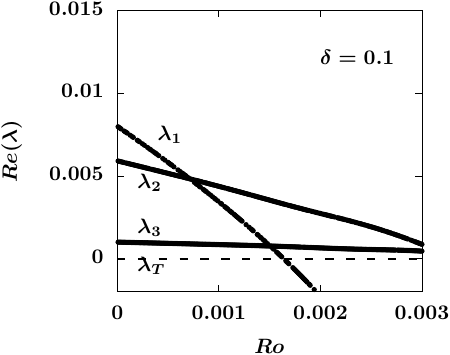} }
\caption{Stability analysis of traveling wave solutions for inner surface flows $(m=-1)$. (a) A typical TW solution (solid) with $Ro = 0.002$, $\delta=0.4$, $\eta=0.01$, $M_0 = 13.124$, $L = 15$, and leading eigenfunctions (dashed curves) associated with eigenvalues $\lambda_1 = 0.0051 \pm 0.03\mathrm{i}$, $\lambda_2=0.0013\pm 0.015\mathrm{i}$, and $\lambda_T = 0$. (b-c) The dependence of the spectrum on system parameters $\delta$ and $Ro$.}	
\label{TWS_ev_inner}
\end{figure}

\begin{figure}
\centering
\subfloat[]{
\includegraphics[width=0.3\textwidth]{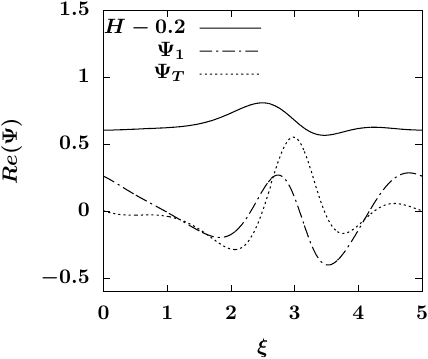} }
\subfloat[]{
\includegraphics[width=0.3\textwidth]{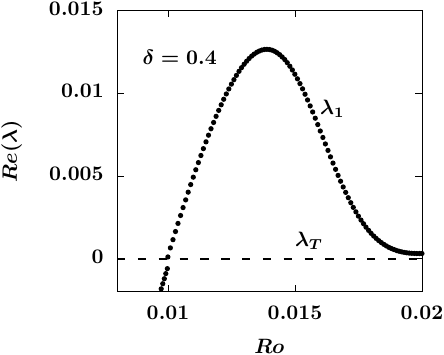} }
\subfloat[]{
\includegraphics[width=0.3\textwidth]{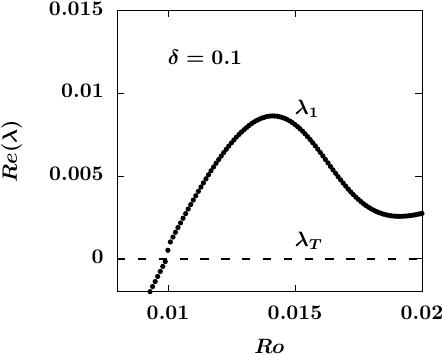} }
\caption{Stability analysis of traveling wave solutions for the outer surface flow $(m=1)$. (a) A typical TW solution (solid) with $Ro = 0.015$, $\delta=0.4$, $\eta=0.01$, $M_0 = 4.569$, $L = 5$ and leading eigenfunctions (dashed curves) associated with eigenvalues $\lambda_1 = 0.01 \pm 0.34\mathrm{i}$ and $\lambda_T = 0$. (b-c) The dependence of the spectrum on $\delta$ and $Ro$.}	
\label{TWS_ev_outer}
\end{figure}
We conduct numerical investigations into the spectrum and corresponding eigenfunctions of the eigenproblem \eqref{eq4.5n}--\eqref{eq4.6n}. In FIG.~\ref{TWS_ev_inner}a,  we present a typical traveling wave solution $H(\xi)$ corresponding to a liquid film flowing along the inner surface of the cylinder ($m=-1$). This slowly-varying TW profile is obtained by numerically solving the nonlinear ODE problem \eqref{eq13aaatw1}-\eqref{eqq_tw2} with $M_0=13.124$, $L = 15$, weak rotation $Ro = 0.002$, $\delta=0.4$, and $\eta = 0.01$. The associated leading unstable eigenvalues of this base state are given by complex conjugate pairs $\lambda_1 = 0.0051 \pm 0.03\mathrm{i}$, $\lambda_2=0.0013\pm 0.015\mathrm{i}$, and a translational eigenvalue $\lambda_T = 0$ that arises from the periodic boundary condition. FIGs.~\ref{TWS_ev_inner}(b-c) show the dependence of the spectrum on the parameters $\delta$ and $Ro$. In both figures, the real parts of the dominant unstable eigenmodes decrease as the rotation becomes stronger, indicating the stabilizing effects of rotation for TW solutions. Comparing FIG.~\ref{TWS_ev_inner}b and FIG.~\ref{TWS_ev_inner}c, we observe that the dominant eigenmodes with a large value of the slip length $\delta$ are greater than those of the case with a smaller slip length. This indicates that the slip effects destabilize the TW solutions for the case $m = -1$.

\par FIG.~\ref{TWS_ev_outer} presents a similar stability study for typical TW solutions for a liquid film flowing down the outer surface of the cylinder, corresponding to the case with $m = 1$. In FIG.~\ref{TWS_ev_outer}a, the slowly-varying TW profile is obtained by solving the ODE problem \eqref{eq13aaatw1}-\eqref{eqq_tw2} with $M_0=4.569, L = 5$, $Ro = 0.015$, $\delta=0.4$, and $\eta = 0.01$. This TW solution has two unstable eigenmodes, a complex conjugate pair $\lambda_1 = 0.01 \pm 0.34\mathrm{i}$ and a translational eigenmode $\lambda_T = 0$, with the associated eigenfunctions presented in dashed curves in FIG.~\ref{TWS_ev_outer}a. 
In this scenario, FIG.~\ref{TWS_ev_outer}(b-c) shows that a positive rotational parameter $Ro > 0.01$ can lead to an unstable TW profile, and its stability can be further amplified by a larger slip length $\delta$. This observation confirms the destabilizing role of rotation ($Ro$) and slip parameters ($\delta$) for TW solutions in the outer surface flow case.

\section{Numerical simulations}\label{sec:6}
To examine the growth of the film instability involving cylinder rotation in the presence of wall slippage, we perform numerical simulations of the PDE (\ref{eq13_model}). Section \ref{sec6a} discusses the choke phenomenon for $S\rightarrow 0$. Section \ref{sec6b} focuses on the breakup or rupture behavior when $S\rightarrow b$. Note that, here $S$ is related to the film thickness $h$ by 
$S(z,t)=b+mh(z,t)$, where $m=\pm1$.

\subsection{Choke phenomenon}\label{sec6a}
In this section, we focus on the ``choke phenomenon'' \cite{camassa2014} or ``plug formation'' \cite{schwitzerlett2023,liu2017} when the thin film flows along the inner surface of the cylinder $(m=-1)$. For the PDE (\ref{eq13_model}), we consider the initial condition as
\begin{subequations}
\begin{equation}\label{choke2}
S(z,0)=R-1+\sum_{j=1}^{J}\mathcal A_j\sin\left(\frac{2\pi}{L}jz\right)+\mathcal B_j\cos\left(\frac{2\pi}{L}jz\right). 
\end{equation}
In (\ref{choke2}), we set $\mathcal A_j=\mathcal B_j=10^{-3}$ and $J=10$, i.e., $20$ harmonic modes. 
\par We approximate the spatial solution by the Fourier method as given below
\begin{equation}\label{choke3}
S(z,t)=\sum_{-(N/2)+1}^{N/2}\widehat S_n\exp\left[\frac{2\pi inz}{L}\right]
\end{equation}
\end{subequations}
In this case, we use adaptive time stepping and set the relative error to $10^{-8}$. 
For the other parameters for investigating the choke phenomenon, we set the other parameters as $\alpha=1/3$, $\eta=0.0625$, and $N=1024$.

\begin{figure}[h!]
\centering
\includegraphics[scale=0.36]{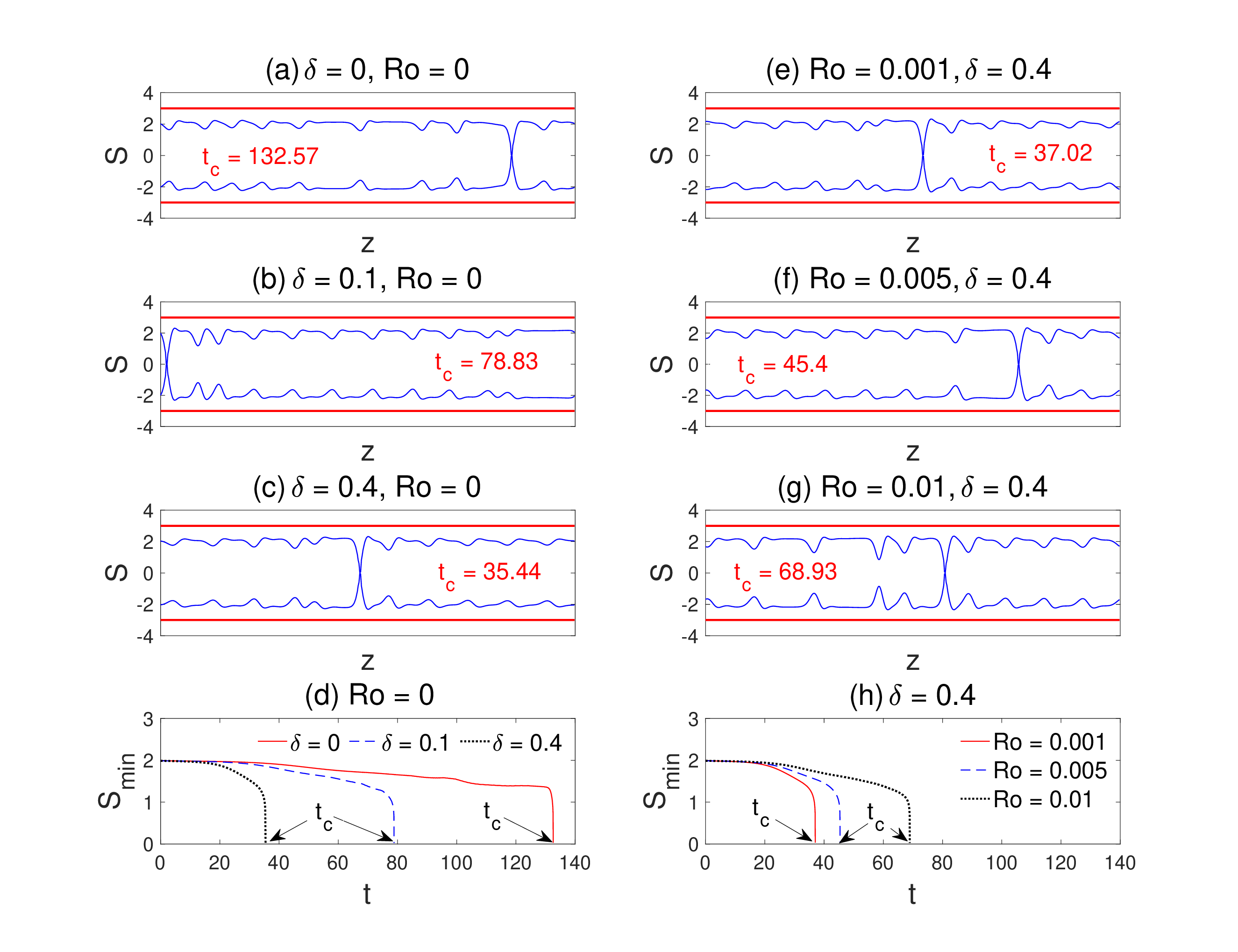}
\caption{(a-c) Interface profiles near the choke time $t = t_c$ for a slippery fixed cylinder $(\delta > 0, Ro = 0)$. (e-g) Interface profiles near choke time $t = t_c$ for a slippery rotating cylinder $(\delta>0, Ro > 0)$. Evolution of the minimum $S(z,t)$ for different (d) slip lengths $\delta$ when $Ro = 0$ and (h) rotation numbers $Ro$ when $\delta = 0.4$. Here, the domain size $L=60$.}
\label{fig7}
\end{figure}
In FIG. \ref{fig7}(a-c), we illustrate the evolution of the free surface when the vertical cylinder is stationary $(Ro=0)$ and the inner cylinder's surface is slippery $(\delta > 0)$. We use two typical values for $\delta$, $\delta = 0.1$ and $\delta = 0.4$, to show the effect of wall slippage on the choke phenomenon. FIG. \ref{fig7}a shows that when $\delta = 0$, indicating no slippage on the inner cylinder surface, the choke time $(t_c)$ is approximately 132.57. In contrast, FIGs. \ref{fig7}b and \ref{fig7}c reveal that the choke time $(t_c)$ decreases as wall slippage $(\delta)$ increases. Additionally, we define the minimum of $S(z,t)$ as $S_{\text{min}}$, i.e., $S_{\text{min}} = \min_z S(z,t)$. In FIG. \ref{fig7}d, $S_{\text{min}}$ is plotted over time for the different slip lengths $\delta$ shown in FIGs. \ref{fig7}(a-c). It is evident that as time progresses, the choke time $t=t_c$ shortens with greater wall slippage. Thus, the left column of FIG. \ref{fig7} demonstrates that wall slip enhances the choking behavior or plug formation. This observation is consistent with Schwitzerlett et al. \cite{schwitzerlett2023}, which also noted that wall slip facilitates plug formation. In FIG. \ref{fig7}(e-g), we explore the impact of rotation on the choke phenomenon when the inner cylinder surface is slippery $(Ro>0,\delta>0)$. The slip length is fixed at $\delta = 0.4$, with other parameters matching those in FIG. \ref{fig7}(a-c). We choose three typical values for $Ro$: 0.001, 0.005, and 0.01. In FIG. \ref{fig7}e, for $Ro = 0.001$, the choke time $(t_c)$ is approximately 37.02, compared to approximately 35.44 in FIG. \ref{fig7}c without rotation. This suggests that rotation increases the choke time $(t_c)$. FIGs. \ref{fig7}f and \ref{fig7}g, where $Ro = 0.005$ and $Ro = 0.01$ respectively, show that the choke time $(t_c)$ further increases with higher $Ro$. In FIG. \ref{fig7}h, $S_{\text{min}}$ is plotted over time for the different rotation numbers $Ro$ shown in FIG. \ref{fig7}(e-g), with the slip length fixed at $\delta=0.4$. Comparing FIGs. \ref{fig7}d and \ref{fig7}h for $\delta=0.4$, it is evident that as time progresses, the choke time $t=t_c$ increases with higher rotation number $Ro$. Therefore, the right column of FIG. \ref{fig7} concludes that while wall slippage promotes choking behavior or plug formation inside a cylinder, introducing rotation can mitigate this effect.

\begin{figure}[h!]
\centering
\includegraphics[scale=0.36]{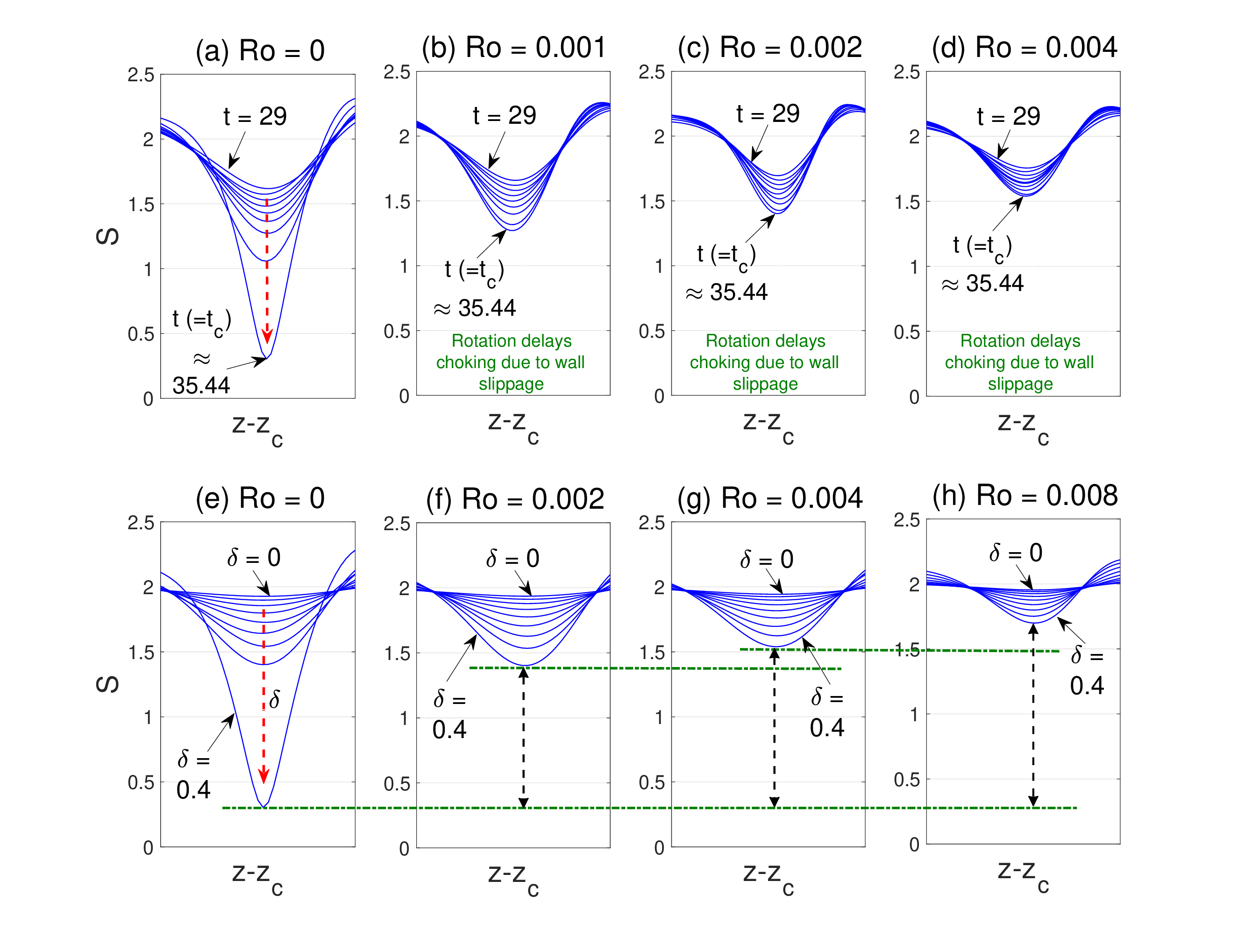}
\caption{(a-d) Influence of $Ro$ on film profiles over time to delay choke behavior with $\delta = 0.4$. (e-h) Influence of rotation number $Ro$ on film profiles at $t \approx 35.44$ to delay choke behavior with slip length $\delta$. Here $L=30$}
\label{fig9}
\end{figure}
In FIG. \ref{fig9}, we depict the delayed choking behavior induced by rotation ($Ro$) in the presence of wall slippage $(\delta)$. In the top panel of FIG. \ref{fig9}, we highlight the free surface configurations to emphasize the delay in choke behavior caused by rotation, with the slip length $\delta$ fixed at 0.4. FIG. \ref{fig9}a shows the surface configuration at time $t = 29$ when $Ro=0$, and we observe that the minimum value of $S$ approaches zero at approximately $t_c \approx 35.44$. Starting from the same initial conditions and varying the rotation number $Ro$ ($Ro = 0.001, 0.002, 0.004$), we present the surface configurations from time $t = 29$ to $t = t_c$ in FIGs. \ref{fig9}(b-d). It is clear that while choke behavior occurs at the critical time $t = t_c$ with $Ro = 0$ for a given wall slippage $\delta=0.4$, the onset of choking is delayed in the presence of rotation ($Ro > 0$), requiring more time for the spatial disturbance to develop. In the bottom panel, we portray the profiles when the slip length varies from $\delta=0$ to $\delta=0.4$ with or without rotation. FIG. \ref{fig9}e shows the free surface configuration without rotation $(Ro=0)$, beginning with $\delta=0$ (no slippage). We observe the minimum value of $S\rightarrow0$ at  $t=t_c\approx35.44$ when $\delta=0.4$. Keeping time constant at $t=t_c$, we keep increasing the rotation number $Ro$ in FIGs. \ref{fig9}(f-h) and show the surface configurations for $\delta$ values ranging from 0 to 0.4. It is evident that at $t_c$, the choke behavior occurs when the cylinder is stationary $(Ro=0)$, but this behavior is delayed as the cylinder begins to rotate $(Ro>0)$. 

\subsection{Breakup behavior and transient solutions}\label{sec6b}
This subsection explores how wall slip and rotation affect the nonlinear evolution of a film flowing on the slippery, rotating vertical cylinder's outer surface $(m=1)$. We mainly focus on the impact of these effects on the breakup behavior and transient dynamic solutions. We set the computational domain as the interval $[-L/2,L/2]$ and impose a periodic condition to simulate the PDE (\ref{eq13_model}). 
\begin{figure}[h!]
\centering
\includegraphics[scale=0.4]{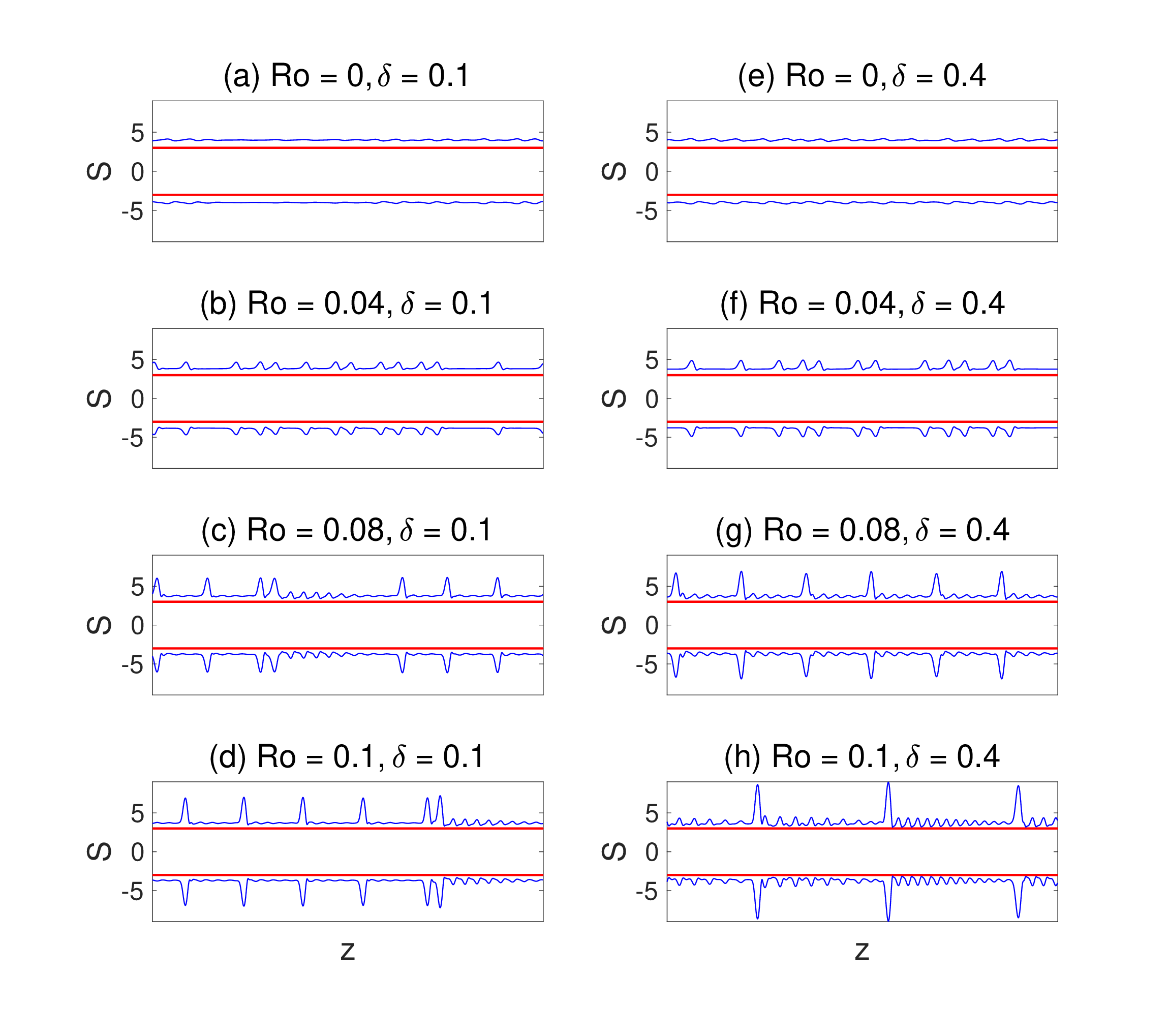}
\caption{Influence of rotation on film profiles when the film flows along the outer surface of a slippery cylinder. The other fixed parameters are $\alpha=1/3$, $\eta=0.0025$ and $L=30$}
\label{fig9brk}
\end{figure}
When a breakup happens for a given parameter, the model collapses at a finite time (say, $t=t_b$), known as a finite-time blow-up. As $t\rightarrow t_b$, the largest interface deformation rapidly grows until it matches the cylinder radius, i.e., when $S(z,t)=b$. At this point, the system reaches a singularity, and solutions cease to exist beyond the time $t=t_b$.
Therefore, detecting the onset of model breakup for a given parameter involves checking whether the interface $S\rightarrow b$. Generally, breakup behavior is local and insensitive to initial and boundary conditions. Therefore, to assess breakup behavior concerning the two key parameters, $\delta$ and $Ro$, we maintain $L=30$ as a fixed value and set the initial condition as 
\begin{equation}
h(z,0)=1-0.2\sin\left(\frac{2\pi}{L}z\right),
\end{equation}
In this case, we have used $512-1024$ Fourier modes (depending on the length of the computational domain). The implicit Gear’s method in time is used, with the relative error set to $10^{-6}$.

\par In FIGs. \ref{fig9brk}(a-d), we illustrate the interface profiles at an instant $t=50$ under the condition where the outer surface of the cylinder is slippery, with a fixed slip length $\delta=0.1$. We consider three representative values to analyze the influence of rotation $Ro$: $Ro=0.04, 0.08,$ and $0.1$. FIG. \ref{fig9brk}a shows the scenario without rotation $(Ro=0)$, where no breakup phenomenon occurs; instead, the interface appears as a quasi-steady traveling wave. In FIG. \ref{fig9brk}b, with $Ro=0.04$, rotational effects cause the interface to break into small droplets separated by varying lengths of inter-drop spacing compared to the non-rotating case (see FIG. \ref{fig9brk}a). In FIG. \ref{fig9brk}c, where $Ro=0.08$, we observe that larger droplets begin to form, and the unstable film between these larger droplets evolves into smaller droplets. Similar time-periodic isolated droplet regime dynamics have been observed in non-rotating fiber coating models \cite{ji2019}.
The droplets are irregularly distributed compared to the $Ro=0.04$ case in FIG. \ref{fig9brk}b. In FIG. \ref{fig9brk}d, where $Ro=0.1$, we find that the big droplets are well developed and larger compared to those in the case of $Ro=0.08$. For this set of parameters, there are almost four small droplets between two large droplets. Additionally, we observe that the smaller droplets are nearly uniform in shape at $Ro=0.04$ (see FIG. \ref{fig9brk}b). However, at $Ro=0.08$ and $Ro=0.1$ (see FIGs. \ref{fig9brk}c and \ref{fig9brk}d), while the larger droplets exhibit almost uniform shapes, the smaller droplets between them do not have a similar structure.

\par FIGs. \ref{fig9brk}(a-d) depict the scenario with a slip length $\delta=0.1$, while FIGs. \ref{fig9brk}(e-f) illustrate the impact of rotation for a higher slip length, $\delta=0.4$, with all other parameters unchanged from FIGs. \ref{fig9brk}(a-d). In FIG. \ref{fig9brk}e, with $Ro=0$ and $\delta=0.4$, the interface forms a more developed quasi-steady traveling wave compared to FIG. \ref{fig9brk}a, without any breakup phenomenon observed. FIG. \ref{fig9brk}f shows film breakup and smaller droplet formation with $Ro=0.04$. The droplets are slightly larger than those in FIG. \ref{fig9brk}b due to the higher slip length. In FIG. \ref{fig9brk}g, for $Ro=0.08$, large droplets form with four smaller droplets between each pair. Unlike the irregular distribution in FIG. \ref{fig9brk}c with $\delta=0.1$, the droplets in FIG. \ref{fig9brk}g for $\delta=0.4$ are almost uniformly spaced. Finally, in FIG. \ref{fig9brk}h, with $Ro=0.1$, more small droplets are observed between each pair of large droplets compared to FIG. \ref{fig9brk}g ($Ro=0.08$), and the large droplets are larger.

\begin{figure}[h!]
\centering
\includegraphics[scale=0.23]{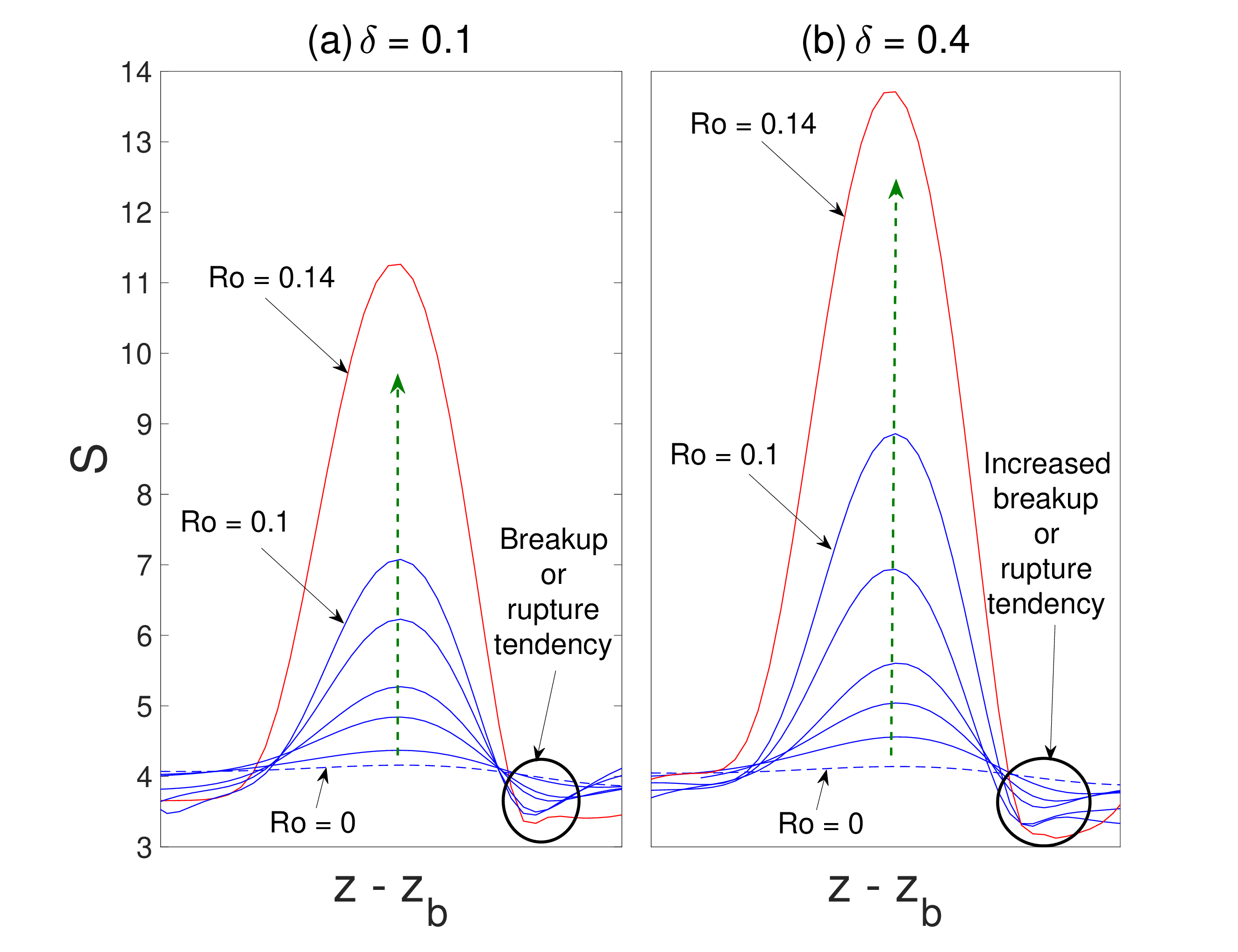}
\caption{Effect of rotation number $Ro$ on film profiles to enhance breakup behavior with slip length. Here $z_b$ is the breakup location, i.e., $\left.z_b=z\right|_{t=t_b}$. The other fixed parameters are $\alpha=1/3$, $\eta=0.0025$ and $L=30$}
\label{fig10brk}
\end{figure}
FIG. \ref{fig10brk} provides a clearer depiction of the liquid-air interfacial profiles, showing how the rupture or breakup phenomena are influenced by both rotation $(Ro)$ and slip length $(\delta)$ for the film flowing along the outer surface of the cylinder. These profiles are plotted at $t=50$. To analyze the effect of $Ro$, we consider the case without rotation $(Ro=0)$ and plot values up to $Ro=0.14$. Both subfigures demonstrate that in the presence of $Ro$, a rupture occurs where the film's minimal thickness approaches zero, and integration rapidly diverges. As $Ro$ increases, this tendency for breakup becomes more pronounced. Additionally, it has been observed that increased slip along the cylinder wall accelerates the pronounced breakup induced by rotation.

\subsection{Effects of rotation and slip on free surface configurations for a Gaussian-shaped profile}\label{sec6c}
In Sections \ref{sec6a} and \ref{sec6b}, we discussed the impact of rotation and wall slip on film profiles using sinusoidal initial conditions. In this section, we explore the effects of a slippery cylinder under rotation on film profiles, considering a Gaussian-shaped initial profile centered at $z=z_0$. 
We perturb the interface as 
\begin{equation}
h(z,0)=1+\widehat\varsigma\exp\left[-0.5\left(z-10\right)^2\right],
\end{equation}
where $\widehat\varsigma$ is a small number that denotes the random disturbance, and $z_0$ is set to 10. In this case, we have used $512-1024$ Fourier modes (depending on the length of the computational domain). Again, the implicit Gear’s method in time is used, with the relative error set to $10^{-6}$.
\begin{figure}[h!]
\centering
\includegraphics[scale=0.42]{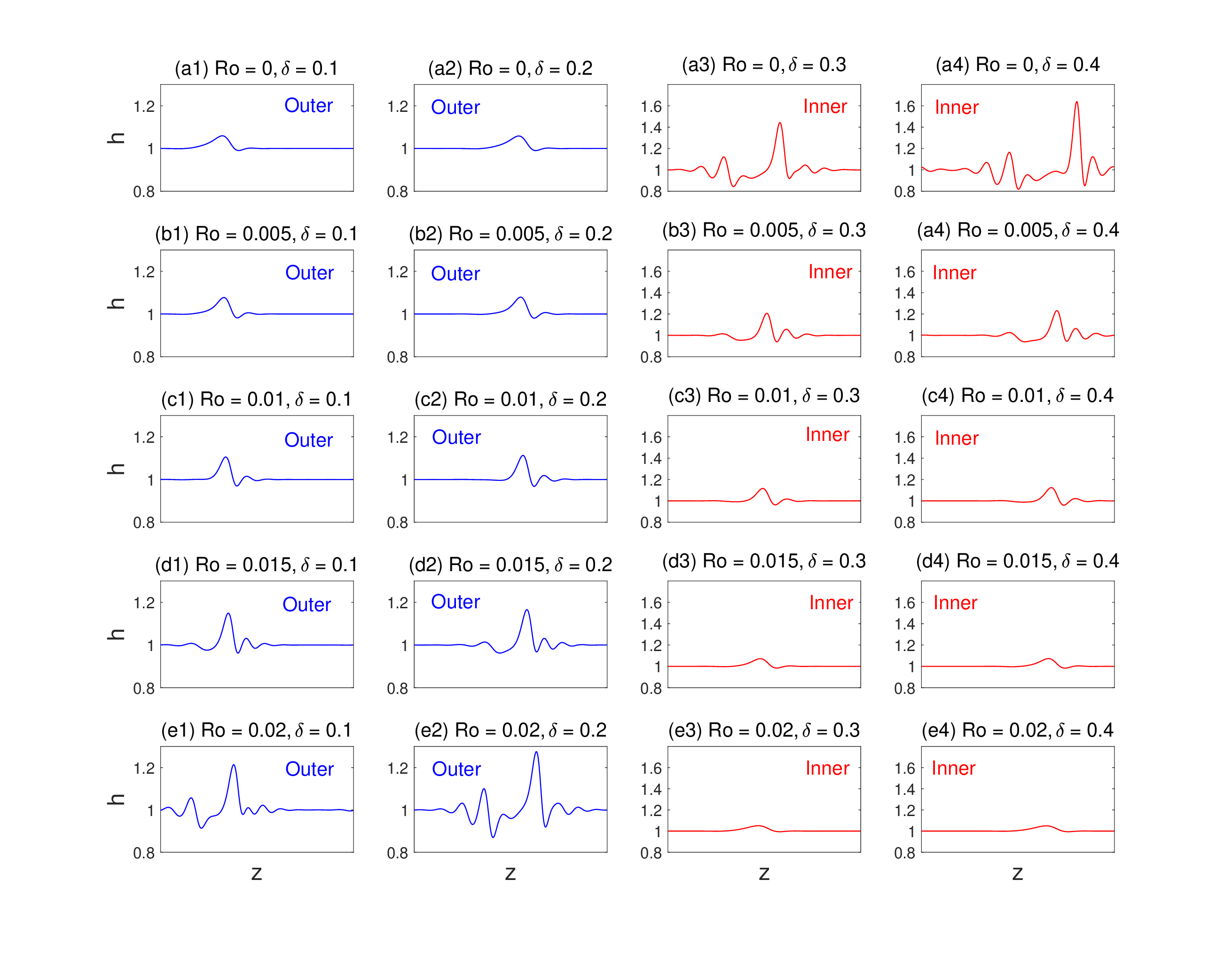}
\caption{Free surface profiles at $t=200$ for the outer surface of the cylinder ($\alpha=1/6$) and the inner surface of the cylinder ($\alpha=1/4$) under the influence of rotation, in the presence of wall slippage. The other fixed parameters are $L=20$, $\eta=0.01$ and $\widehat\varsigma=0.1$}
\label{fig10}
\end{figure}
\par FIG. \ref{fig10} presents the free surface configurations for a rotating slippery cylinder $(\delta > 0, Ro > 0)$. The first two columns illustrate the film flow along the outer surface of the cylinder, while the third and fourth columns depict the flow along the inner surface. In the first column, with a fixed slip length of $\delta = 0.1$, profiles for $Ro = 0, 0.005, 0.01, 0.015,$ and $0.02$ are shown. This column reveals that as the rotation rate increases, the interfacial waves form higher humps, indicating increased instability compared to the non-rotating case ($Ro = 0$) displayed in FIG. \ref{fig10}a1. Additionally, as $Ro$ increases, the one-hump solitary-like wave structure becomes more oscillatory, as seen in FIG. \ref{fig10}e1. In the second column, the slip length is increased to $\delta = 0.2$, with other parameters remaining the same. Comparing the first and second columns, it is evident that higher wall slippage results in more pronounced surface distortions. In the third column, the slip length is set to $\delta=0.3$ for the inner surface flow of the cylinder. This column shows that with increased rotation, the interfacial waves exhibit lower humps, thereby reducing instability compared to the non-rotating case ($Ro = 0$) shown in FIG. \ref{fig10}a3. Comparing FIGs. \ref{fig10}a3 and \ref{fig10}e3, it can be observed that with higher $Ro$, the highly distorted liquid-gas interface transitions into a one-hump solitary-like wave structure. In the fourth column, the slip length is further increased to $\delta=0.4$. Here, wall slippage amplifies the disturbances; however, higher rotation mitigates these disturbances.

\begin{figure}[h!]
\centering
\includegraphics[scale=0.28]{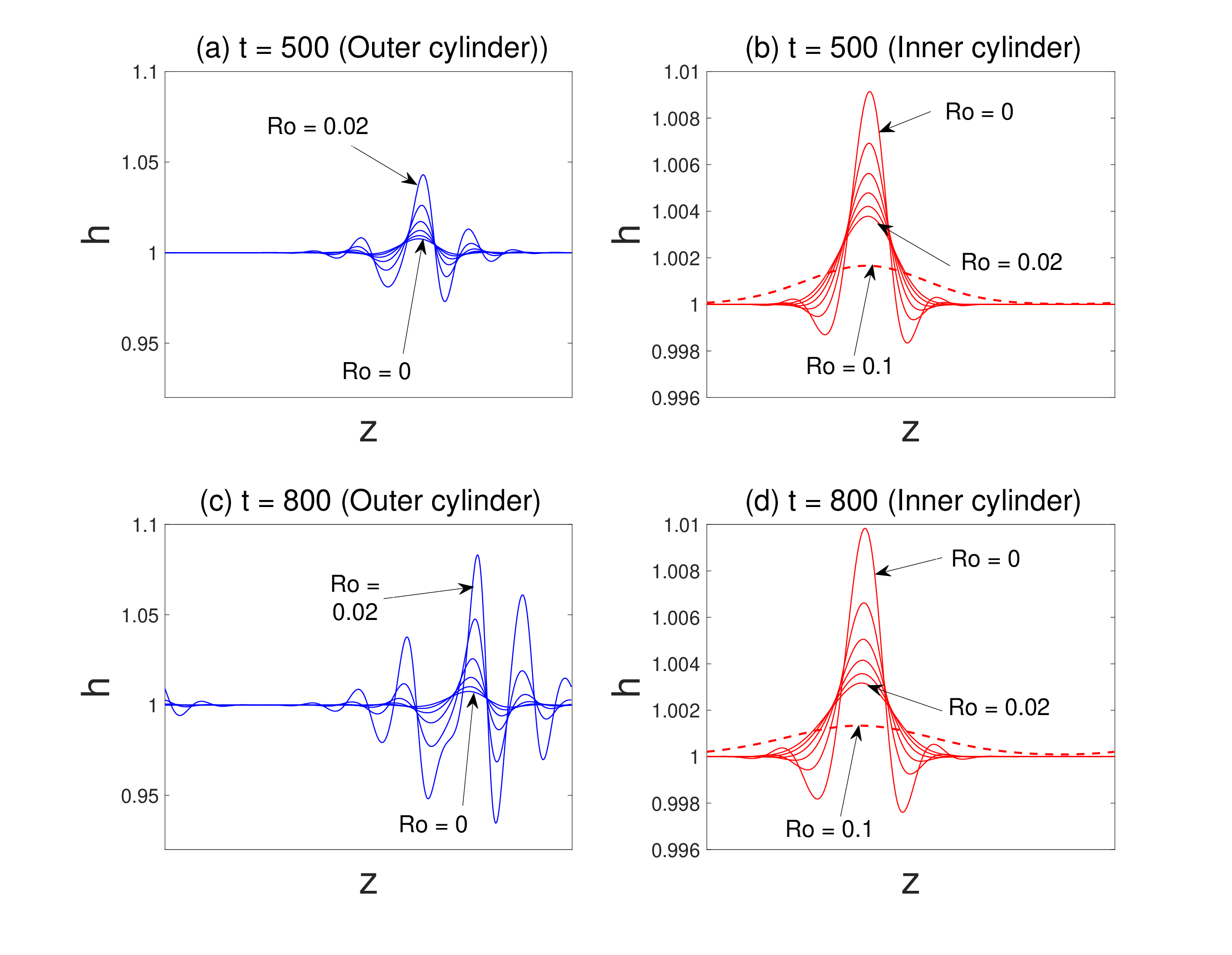}
\caption{Free surface profiles at two different instants for the outer ($m=1$) and inner ($m=-1$) surfaces of the slippery cylinder under the influence of rotation. The other fixed parameters are $\alpha=1/6$, $\delta=0.2$, $\eta=0.04$, $L=40$, and $\widehat\varsigma=0.01$}
\label{fig11}
\end{figure}
In FIG. \ref{fig10}, we present the interfacial shape at $t=200$. In FIG. \ref{fig11}, we show how the film profiles are influenced by rotation ($Ro$) at longer times, $t=500$ and $t=800$. The left panel illustrates the film flow along the outer surface of the slippery cylinder, while the right panel shows the flow along the inner surface. The slip length $\delta$ is fixed at 0.2, and the computational domain length is $L=40$ for both outer and inner cylinder scenarios. FIGs. \ref{fig11}a and \ref{fig11}c indicate that at later times, oscillations become more significant with higher rotation when the film flows along the outer surface of the slippery cylinder. Conversely, FIGs. \ref{fig11}b and \ref{fig11}d demonstrate that although surface wave instability slightly amplifies over time, there is no significant development of an oscillatory wave structure for $Ro=0$. Even at longer times, rotation reduces surface wave instability in the presence of wall slippage. Additionally, for FIGs. \ref{fig11}b and \ref{fig11}d, the free surface profiles are plotted at $Ro=0.1$. At this $Ro$ value, the crest is almost reduced, indicating a high stabilizing effect of rotation on the inner surface flow, even when the cylinder wall is slippery.

\section{Summary and conclusions}\label{sec:7}
In this study, we have conducted a detailed theoretical investigation of gravity-driven viscous film flows along a vertical cylinder with a slippery surface. To account the effect of rotation on the flow dynamics and stability, we have considered the vertical cylinder to be rotating around its vertical axis. The slipperiness of the cylinder wall is accounted for by employing the Navier-slip boundary condition (velocity at the wall is assumed to vary proportionally to the normal derivative of velocity at the solid-liquid interface). To describe the dynamics of the film interface, we have assumed that the mean thickness of the film is much smaller than its characteristic length in the axial direction. Our model captures the coupled effect of rotation and wall slippage for the flow along the outer and inner surface of the vertical cylinder.

\par We have performed a linear stability analysis to investigate the impact of rotation on the stability of axisymmetric disturbances in flows where wall slip occurs on both the inner and outer surfaces. Our findings indicate that while wall slip enhances flow instability on both surfaces, rotation stabilizes the inner surface but destabilizes the outer surface. In addition to the linear analysis, we have conducted a weakly nonlinear stability analysis to explore how nonlinear effects influence the stability thresholds in our model. This investigation revealed the existence of both supercritical and subcritical regimes, influenced by both wall slip and rotation for flows on both surfaces. Furthermore, we have extended our study to consider traveling wave solutions for both outer and inner surface flows, examining how rotation modifies their characteristics and velocities in the presence of wall slip. We have also evaluated the stability of these traveling wave solutions under conditions where the cylinder rotates, and the wall exhibits slip. Our results demonstrate that incorporating rotational effects can partially or completely mitigate interfacial instabilities in traveling waves for the inner surface flow within a fixed domain. Conversely, for the outer surface flow, rotation tends to exacerbate interfacial instabilities in traveling waves. Finally, numerically simulating the evolution equation, we demonstrate that introducing rotation effectively delays the onset of choking phenomena in the inner surface flow, even when wall slippage is present. Conversely, our simulations also reveal that rotation amplifies the breakup process in the system for flow along the outer surface of the cylinder, particularly pronounced in the presence of slip.

\par This study holds potentially significant implications for industrial coating processes and the lubrication of pipes with Newtonian fluids on slippery solid surfaces. Our analysis reveals that rotation can mitigate slip-induced instability for inner surface flow but not for outer surface flow. Our analysis, relevant to fluids with negligible inertia, could inspire future research to experimentally validate these findings and further explore thin-film flows along slippery cylinders at low to moderate Reynolds numbers. Future investigations will include evaluating non-isothermal effects and examining non-Newtonian films on similar geometries.

\section*{Conflict of interest}
The authors have no conflicts to disclose.

\section*{DATA AVAILABLITY STATEMENT}
The data that supports the findings of this study are available within the article.

\section*{Acknowledgements}
AKG acknowledges the support from Science and Engineering Research Board (SERB) grant SIR/2022/001357, Government of India. HJ acknowledges support from grant NSF DMS 2309774.

\section*{Declaration of Generative AI and AI-assisted technologies in the writing process}
During the preparation of this work, the authors used ChatGPT in order to improve the language and readability of the article. After using this tool, the authors reviewed and edited the content as needed and take full responsibility for the content of the publication.

\end{document}